\documentclass{pasj00}

\Received{$\langle$$\rangle$}
\Accepted{$\langle$$\rangle$}
\Published{$\langle$publication date$\rangle$}
\SetRunningHead{Astronomical Society of Japan}{Usage of \texttt{pasj00.cls}}

\usepackage{times}
\usepackage{lineno}
\usepackage{color}
\usepackage{ifthen}

\newcommand{\fx}{\ensuremath{F_{\rm{X}}}}
\newcommand{\fuv}{\ensuremath{F_{\rm{UV}}}}
\newcommand{\fdisk}{\ensuremath{F_{\rm{disk}}}}
\newcommand{\nh}{\ensuremath{N_\mathrm{H}}}
\newcommand{\ledd}{\ensuremath{L_\mathrm{edd}}}

\newcommand{\ergcms}{\mathrm{erg}\ \mathrm{cm}^{-2}\ \mathrm{s}^{-1}}
\newcommand{\ergcmse}{\times{10}^{-8}\ \mathrm{erg}\ \mathrm{cm}^{-2}\ \mathrm{s}^{-1}}

\newcommand{\Tin}{T_\mathrm{in}}
\newcommand{\rin}{r_\mathrm{in}}
\newcommand{\Rin}{R_\mathrm{in}}

\newcommand{\msolar}{M_{\odot}}

\usepackage{xcolor}
\usepackage{changebar}
\definecolor{ins}{rgb}{0.5, 0, 0}
\definecolor{del}{rgb}{0, 0, 0.5}

\newcommand{\xspec}{\tt{xspec}}
\newcommand{\ftools}{\tt{ftools}}
\newcommand{{\maxi}}{MAXI}
\newcommand{{\swift}}{Swift}

\begin{document}

\SetRunningHead{Nakahira et al.}{ Study of mass accretion in MAXI~J1910$-$057}

\title{A new X-ray nova MAXI~J1910$-$057 ($=$ Swift~J1910.2$-$0546)\\ 
and mass-accretion inflow 
}
\author{
Satoshi \textsc{Nakahira}\altaffilmark{1}, 
Hitoshi \textsc{Negoro}\altaffilmark{2}, 
Megumi \textsc{Shidatsu}\altaffilmark{3}, 
Yoshihiro \textsc{Ueda}\altaffilmark{3} ,\\
Tatehiro \textsc{Mihara}\altaffilmark{4}, 
Mutsumi \textsc{Sugizaki}\altaffilmark{4}, 
Masaru \textsc{Matsuoka}\altaffilmark{4} and 
Takuya \textsc{Onodera}\altaffilmark{2}
	}
\altaffiltext{1}{ISS Science Project Office, Institute of Space and Astronautical Science (ISAS), Japan Aerospace Exploration Agency (JAXA), 2-1-1 Sengen, Tsukuba, Ibaraki 305-8505}
\altaffiltext{2}{Department of Physics, Nihon University, 1-8-14, Kanda-Surugadai, Chiyoda-ku, Tokyo 101-8308}
\altaffiltext{3}{Department of Astronomy, Kyoto University, Oiwake-cho, Sakyo-ku, Kyoto 606-8502}
\altaffiltext{4}{MAXI team, Institute of Physical and Chemical Research (RIKEN), 2-1 Hirosawa, Wako, Saitama 351-0198}

%


\maketitle

\begin{abstract}

We report on a long-term monitoring of a newly discovered X-ray nova,
MAXI~J1910$-$057 ($=$ Swift~J1910.2$-$0546), by {\maxi} and {\swift}. 
The new X-ray transient was first detected on 2012 May 31 by
{\maxi} Gas Slit Camera (GSC) and {\swift} Burst Alert Telescope (BAT)
almost simultaneously.  We analyzed X-ray and UV data for 270 days
since the outburst onset taken by repeated {\maxi} scans 
and {\swift} pointing observations. The obtained X-ray light curve
for the inital 90 days is roughly represented by a fast-rise and
exponential-decay profile. 
However, it re-brightened on the $\sim 110$ days after the onset and 
finally went down below both GSC and BAT detection limits on the 240 day.
All the X-ray energy spectra are fitted well with a model
consisting of a multi-color-disk blackbody and its Comptonized hard
tail. During the soft-state periods,
the inner-disk radius of the best-fit model were almost constant. 
If the radius represents the innermost stable circular orbit
of a non-spinning black hole and the soft-to-hard transitions occur at
1--4\% of the Eddington luminosity, the mass of the compact object is
estimated to be $>2.9 \msolar$ and the distance to be $>1.70$ kpc.
The inner-disk radius became larger in the hard / hard-intermediate
state. This suggests that the accretion disk would be truncated.
We detected an excess of the UV flux over the disk blackbody component
extrapolated from the X-ray data, which can be modelled as reprocessed
emission irradiated by the inner disk.
We also found that the UV light curve mostly traced the X-ray
curve, but a short dipping event was observed in both the UV and the X-ray bands
with a 3.5-day X-ray time lag.
This can be interpreted as the radial inflow of accreting matter from the outer UV region to the
inner X-ray region.

\end {abstract}

\section{Introduction}

Galactic low-mass X-ray binaries (LMXBs) are classified into two
different types, neutron star binary and black hole X-ray binary
(BHXB).  While some of LMXBs are persistently on, most of the BHXBs
are thought to be transients that are normally in a quiescent state
and occasionally exhibit episodic brightening called ``outburst''.
The outburst is triggered by sudden increase of mass accretion rate
onto the compact star.
There are two competing models that regulate the accretion flow, namely,
``mass transfer instability model'' \citep{MTI} and ``disk thermal
instability model'' \citep{mineDTI}.

During the outburst, BHXBs change its X-ray luminosity in several
orders of magnitude and transit through two major spectral states, 
``soft state'' and ``hard state''.  The X-ray spectrum in the soft
state (hereafrer SS) is characterized by an ultra-soft component
which originates from optically-thick and geometrically-thin accretion
disk, called ``standard disk'' \citep{stddisk}.  Many observations
show that the inner-most radius $\rin$ remains constant 
irrespective of variation of the inner most 
temperature. Therefore, the $\rin$ is believed to reflect the inner
stable circular orbit (ISCO).  The hard state (hereafrer HS) is
approximated by a single power-law with a photon index 1.4--2.0 and a
high-energy cutoff at $\sim$100 keV.
\citet{sunyaev_comp} successfully explained this emission by thermal Comptonization 
of soft  seed photons in the hot corona. 
The seed photons are believed to come from the standard disk which is
truncated in a large $\rin$ (\cite{truncate}, \cite{cygx1_makishima}).
The hysteresis in X-ray intensity between the
HS-to-SS and SS-to-HS state transitions produces a q-shaped track in a
hardness intensity diagram, called ``q-curve'' \citep{qcurve}.

Past X-ray observations of BHXBs including {\it RXTE} and {\maxi}
revealed a variety of light-curve profiles among BHXBs
(e.g. \cite{novalc_chen}). ``X-ray nova'' is a subclass of these BHXB
transients, which includes A0620$-$00 \citep{0620}, GS\
1124$-$68 \citep{1124}, GS 2000$+$25 \citep{2000} and GRO
J0422$+$32 \citep{0422}. They are characterized by a
simple light-curve profile of fast-rise and exponential-decay normally followed
by late {\it re-flaring} \citep{xraynovae}. The simple profile allows
us to study accretion inflow of the black-hole binary system apart
from complicated behavior around the inner disk.

A newly discovered X-ray transient MAXI~J1910$-$057 ($=$
Swift~J1910.2$-$0546) is categorized into X-ray novae from the profile 
of the X-ray light curve.  The source was detected by Gas-Slit Camera
(GSC; \cite{gscmihara}) onboard Monitor All-sky X-ray Image ({\it
MAXI}; \cite{maxi}) operated on the international space station (ISS).
On 2012 May 31 22:36 (MJD=56078.941667), {\maxi} transient
alert system (NovaSearch; \cite{novaalert}) detected an enhanced
X-ray emission at (R.A., Dec)$=$(\timeform{287D.8}, \timeform{-5D.8}), 
and then issued a flash report tagged by Trigger ID=6078777285
to the ``New-transient'' alert mailing
list\footnote{https://maxi.riken.jp/mailman/listinfo/new-transient}
maintained by {\maxi} team. The updated information was also
reported to the Astronomer's Telegram \citep{1910maxiatel1}.  Swift
Burst Alert Telescope (BAT; \cite{swiftbat}) independently detected
the source \citep{1910swiftatel1}. Therefore, this source is also
referred to as Swift~J1910.2$-$0546.  The first {\swift} pointing
observation, performed at 2012 June 1 14:41 (MJD 56079.611806), 
refined the source position to (R.A., Dec)
(J2000)~$=$~(\timeform{19h10m22s.78}, \timeform{-05D47'58.0''})
(\cite{1910swiftatel2}) with an uncertainty of  \timeform{3.5''} radius. 
The source is also identified with optical observations \citep{1910grondatel}.
Previous works on this source
by \citet{1910reis} presented results of two {\it XMM--Newton}
observations in a hard or hard-intermediate state, and they reported
a possibility of retrograde spin.
\citet{1910Degenaar} investigated the correlation around the re-flare 
using multi wavelength light curves from near-infrared to X-ray bands.

In this paper we studied this ``classical'' X-ray nova from the onset
on 2012 May 31 (MJD=56078) to the end of the activity around 
2013 January 26 (MJD=56318) using data of{\maxi} and {\swift}, complementally.  
Hereafter, $t$ denotes the elapsed time since the date of the onset, MJD$=$56078.0. 
We performed light-curve and spectral analyses of X-ray and UV data
utilizing {HEASOFT} version 6.13 with {\xspec} version 12.8.0m,
and {\maxi} specific tools developed on the basis of {\ftools}.
Errors represent 1$\sigma$ and 90\% confidence limits for light curve
data and best-fit spectral parameters, respectively, throughout the
following sections.

\section{Instruments and Data Reduction}

\subsection{{\maxi}}

{\maxi}, in operation since 2009 August, has two scientific
instruments, GSC and SSC (Solid-state Slit
Camera; \cite{ssctomida}). Both GSC and SSC cover two
instantaneous rectangular fields of view (FOV), aimed at the earth
horizon and the zenith directions.  Each FOV of 3$\times$160 deg$^2$
in GSC and 3$\times$90 deg$^2$ in SSC is determined by a slit and slats
collimators. The target visibility is determined by an ISS
attitude which rotates with the orbital motion.  Each visible time
(hereafter referred to as ``transit'') lasts for 50--150 sec for GSC,
and 40--60 sec for SSC.  The effective area changes due to the
triangular-shaped collimator transmission with a peak value of 4--5
and 0.6--0.9 cm$^2$ for GSC and SSC, respectively.

\subsubsection{GSC data}

We used {\maxi} specific analysis tools and the script which are
developed by the {\maxi} team.  These software are utilized in the
{\maxi} on-demand data processing system accessed viea web
interface\footnote{http://maxi.riken.jp/mxondem}\citep{nkhrjssij},
where the latest calibration information is applied.  In the period 
of interest in this paper, six GSC counters out of the twelve were turned on,
and others were kept off for redundancy \citep{gscsugizaki}.  We
utilized data of GSC \#0, 2, 7 operated at the high voltage of 1550 V and
of GSC \#4, 5 at 1650 V.  Here we did not utilize GSC\#3 because it was
degraded and not suitable for spectral analysis.  The X-ray event data
archived as the GSC event revision 1.4 was used.  Then we set the
source regions and the background regions for the analysis.  For
source regions, a circle centered at the target position was used.
Because 1550 V counters have relatively wider PSF than those of 1650 V, we use
the radius r$_{s}$ of \timeform{2D.0} and \timeform{1D.5},
respectively.  For background regions we used an annulus region
between an inner radius of r$_s$+\timeform{0D.1} and an outer radius
of r$_s$+\timeform{1D.0} centered on the target position.
Additionally we excluded X-ray events within \timeform{1D.5} from an
X-ray burster 4U~1916--053 which locates
\timeform{2D.1} apart from the target source.
We discarded these data where the source and background regions 
1) were shaded  by solar panels or other ISS structure, or
2) the scan of either region was not completed.
As a result, we collected 2886 transits of 493.6 cm$^{2}$ ksec.

\subsubsection{SSC data}

The SSC observation was conducted in similar manner as GSC. Hence, SSC
data is analyzed in the similar manner as the GSC except for the data
screening.  We first removed the hot pixels with {\tt cleansis} in
{\ftools}. We then collected single-pixel events (G0) in the energy
band below 1.84 keV, and included split events (G12) above 1.84 keV.
The energy gain was corrected by using copper K$_\alpha$ emission line
from the SSC camera body, and the gain temperature dependence was also
corrected.  Since the SSC data were contaminated by the
visible/infrared light from the Sun and Moon \citep{ssctsumemi}, we
selected data taken during the ISS night time, and then eliminated
such events that the angular separation from the Moon is $<5^\circ$.
In this manner, we collected 844 transits of 22.2 cm$^{2}$ ks.  To
avoid thermal noise due to relatively higher CCD temperature, we
ignored energy bin below 0.7 keV, above 7 keV dominated by background.
Since the response function of SSC has calibration uncertainties of
about 10\% in normalization, we performed simultaneous model fits to
both GSC and SSC spectral data introducing a cross normalization
factor.  We set the factor equal to 1.0 in GSC, and let free in SSC.

\subsection{Swift}

Since the discovery of the new X-ray source, MAXI~J1910$-$057 ($=$
Swift~J1910.2$-$0546), {\swift} \citep{swiftxrt} pointing
observations were carried out with X-Ray Telescope (XRT) and UV
Optical Telescope (UVOT; \cite{UVOT}) every one to $\sim$ten day
interval.  Due to the Sun angle constraint, no observations were
available from 2012 November 26 to 2013 March 9.  We utilized all the
observation data taken until 2012 November, whose target IDs are
32480, 32521 and 32742.

\subsubsection{XRT data}
We utilized both data taken with PC and WT modes with an
exposure of a few ks.  All X-ray spectra and light-curve files
were produced on the data-archive web site supplied by the UK {\swift} Science
Data Centre at the University of Leicester
\citep{swiftukwebana}.  We performed spectral analysis
with a response matrix file {\tt swxwt0to2s6\_20010101v014.rmf} in the
0.4--10 keV energy band.

\subsubsection{UVOT data}

We collected UVOT data from NASA/HEASARC data center.
We extracted fluxes and energy spectra with {\tt uvotproduct} and
{\tt uvot2pha} in {\ftools}, respectively.  Since observations of
ID$=$32521 were performed with a single UVM2 filter (2246 $\AA$), the
light curve analysis is focused on the UVM2 band.
We integrated image using {\tt uvotimsum}, and then
estimated background carefully from the source free region.

\subsubsection{BAT light curve}

The Burst Alert Telescope (BAT) on {\swift} is an all-sky monitor
to search for Gamma-ray bursts. While searching for bursts, it is
capable of monitoring hard X-ray sources.  
We utilized the archived 15--50 keV light-curve data with a one-day
time bin, available on the ``BAT Transient Monitor'' website
(\cite{battransient})
\footnote{http://swift.gsfc.nasa.gov/docs/swift/results/transients}.

\section{Analysis results}

\subsection{Light curves and hardness variations}\label{results_lc}

\begin{figure}
  \begin{center}
    \FigureFile(85mm,85mm){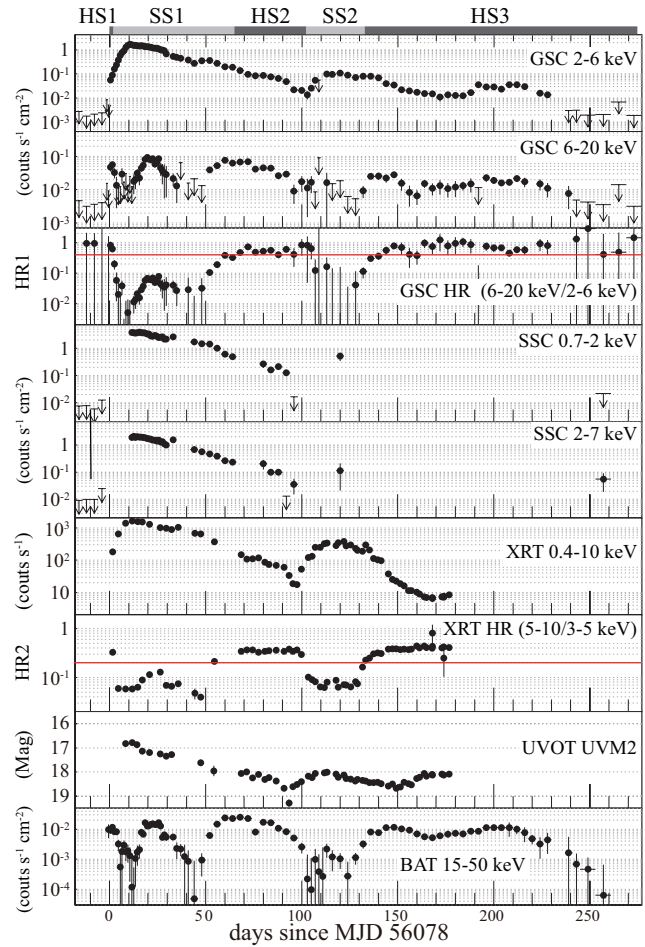}
  \end{center}
  \caption{
  Upper three panels are {\maxi}/GSC 2--6 keV, 6--20 keV counts
  and hardness ratio between 6-20 keV and 2-6 keV band (HR1).
  The next six panels show {\maxi}/SSC 0.7--2 keV and 2--7 keV lights curve, 
  then {\swift}/XRT 0.4--10 keV, 5--10/3--5 keV (HR2), 
  {\swift}/XRT light curve of UVM filter, 
  and {\swift}/BAT 15-50 keV.
  Five periods split by {\maxi}/GSC HR=0.4 and {\swift} HR=0.2 (red lines) 
  are shown at the top of the figure.
  }
  \label{fig:LC}
\end{figure}

In figure \ref{fig:LC}, we plot GSC 2--6 keV and 6--20
keV light curves and their hardness ratio (HR1), 
SSC 0.7--2 keV and 2--7 keV light curves,
XRT 0.4--10 keV light curve and hardness ratio of the 5--10 keV to the
3--5 keV (HR2), UVOT light curve with UVM2 filter, and BAT 15-50 keV
light curve.  Data are binned by 0.5, 1, 2, 4 or 8 days depending on
photon statistics of each bin.
GSC scanned the source almost every day
since the mission started in 2009 August.  We investigated the GSC event data 
before the outburst onset. However, no significant excess 
over the background was detected. 
The hardness variation implies that the source
exhibited spectral transition between HS and SS. Hence, 
we divided the entire periods into 
five intervals, labeled as 
HS1 ($t<$ 3 d), 
SS1 (3 d $<t<$ 66 d), 
HS2 (66 d $<t<$ 102 d), 
SS2 (102 d $<t<$ 132 d), 
and HS3 (132 d $<t$) 
in figure \ref{fig:LC},
according to the HR1 below/above $0.4$.
or HR2 below/above 0.2.  
We employed HR1 in $t<80$ d and HR2 in $t>80$ d
according to the data frequency and the photon statistics.

As noticed from the GSC light curves, the 2--6 keV flux increased rapidly
in the outburst beginning and reached the peak of 1.70 counts
cm$^{-2}$ s$^{-1}$ $\simeq$ 950 mCrab at $t\sim 10$ d, where the 6--20
keV flux decreased.  This caused the state transition from HS1 to SS1.
While the X-ray emission in the soft state is dominated by a soft component, 
a weak hard X-ray peak was also seen at around $t=15-30$ d \citep{kimuraatel}.  
In the SS1, the 2--6 keV flux declined steady, but the 6--20 keV
turned to increase at $t=50$ d, and then the source returned
to the hard state (HS2) \citep{nkhratel}.

After $t\sim 100$ d, the source repearted the same state-change cycle
of the hard(HS2)-soft(SS2)-hard(HS3) while the intensity declined
gradually.
In the HS3, the X-ray intensity varied to some degree
keeping a hard spectrum.  Further decay began around $t=230$ d, and
gone below both GSC and BAT detection limits after $t=240$ d.

The UV light curve showed similar shape to
soft X-ray light curve from SS1 until $t=$ 140 d.  However, it showed
a clear deviation after then.  
The correlation between UV and X-ray is studied in section \ref{x-uv_corr}.

The entire structures of light curve is
characterized with a fast rise and an exponential decay, followed by a
re-flaring and latter long tail.  This feature resembles the classical
black hole X-ray novae \citep{xraynovae}.

\subsection{Evolution of X-ray Spectrum}\label{xray_spec}

\begin{figure*}
  \begin{center}
    \FigureFile(55mm, 55mm){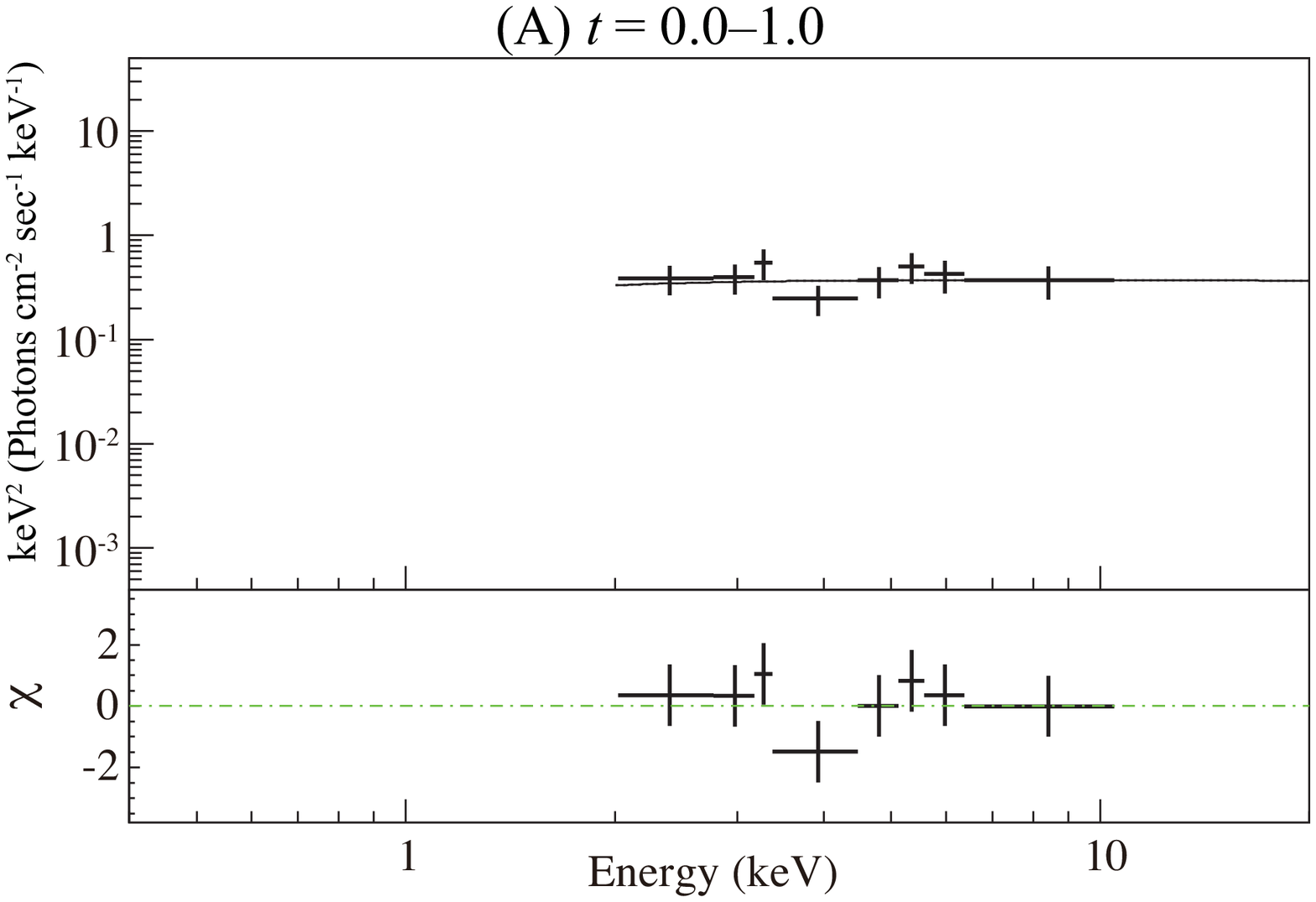}
    \FigureFile(55mm, 55mm){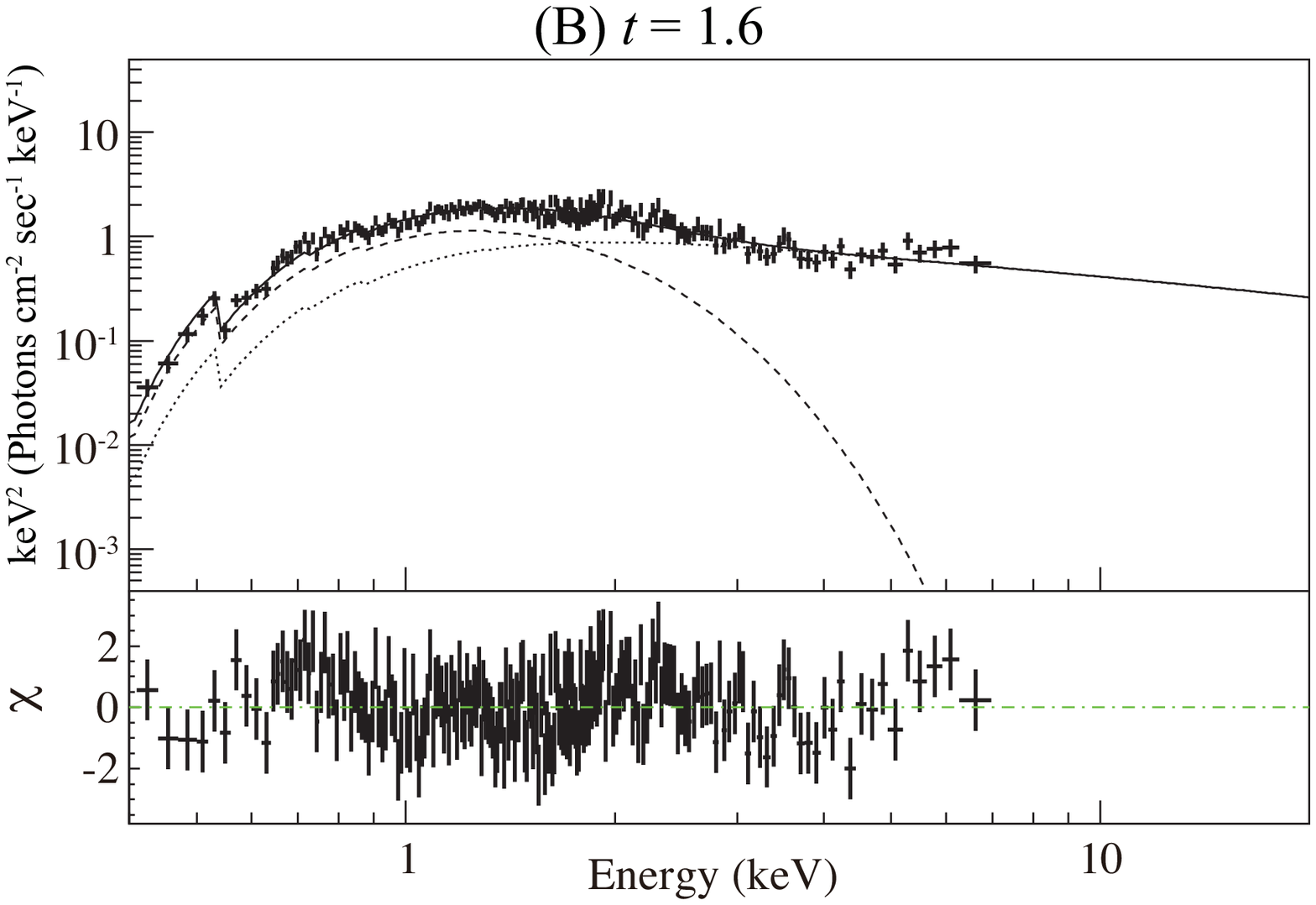}
    \FigureFile(55mm, 55mm){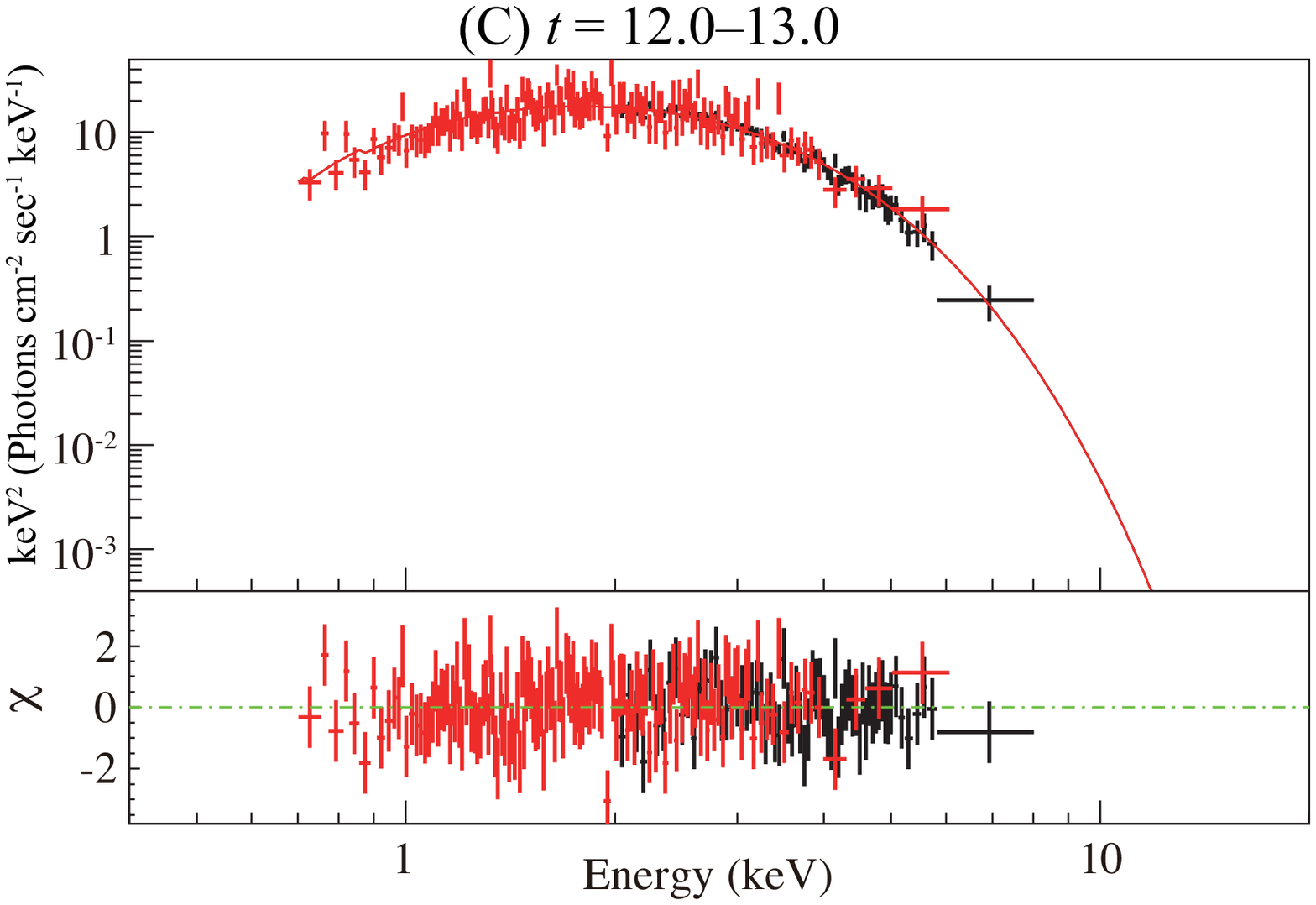}
    \FigureFile(55mm, 55mm){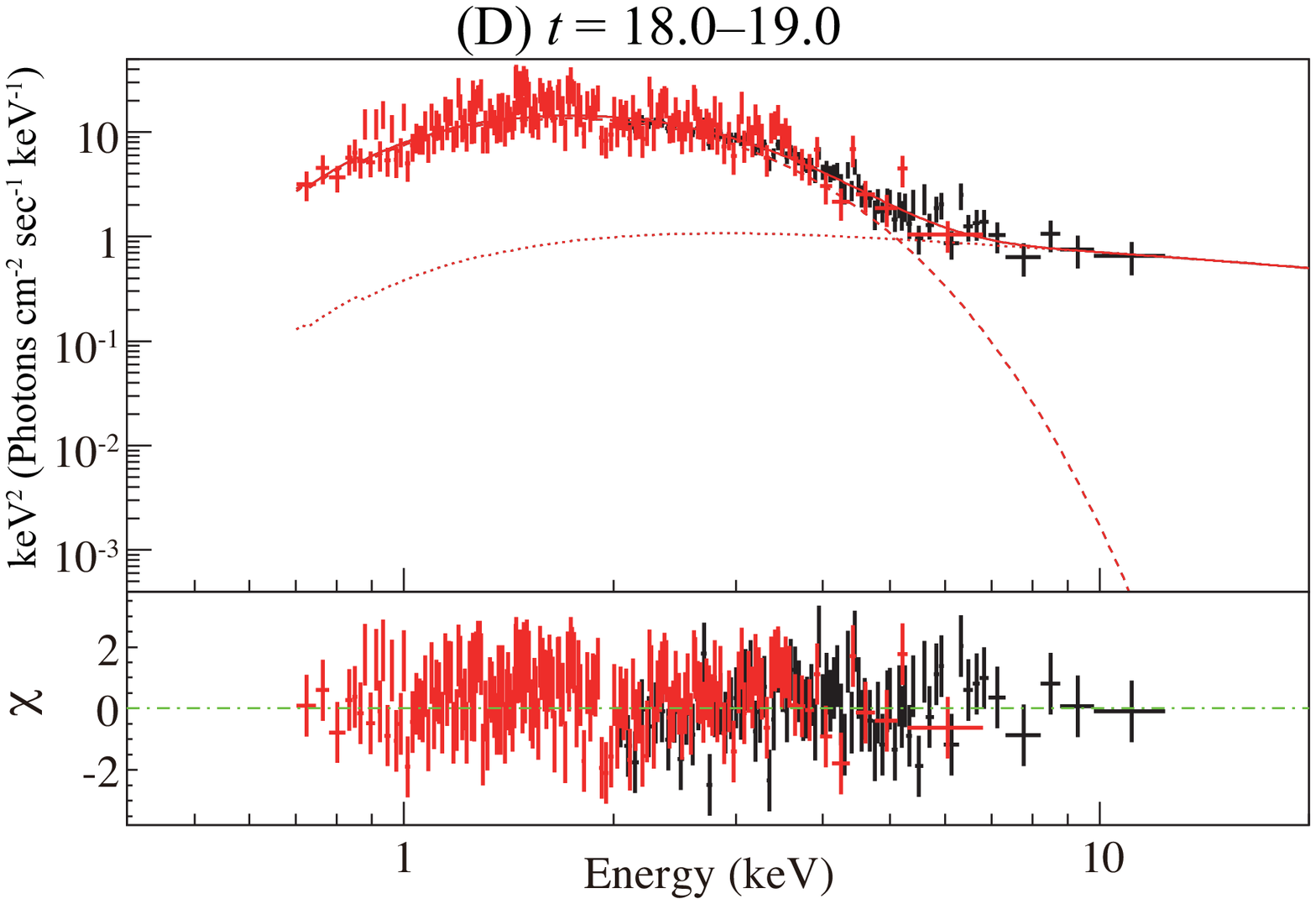}
    \FigureFile(55mm, 55mm){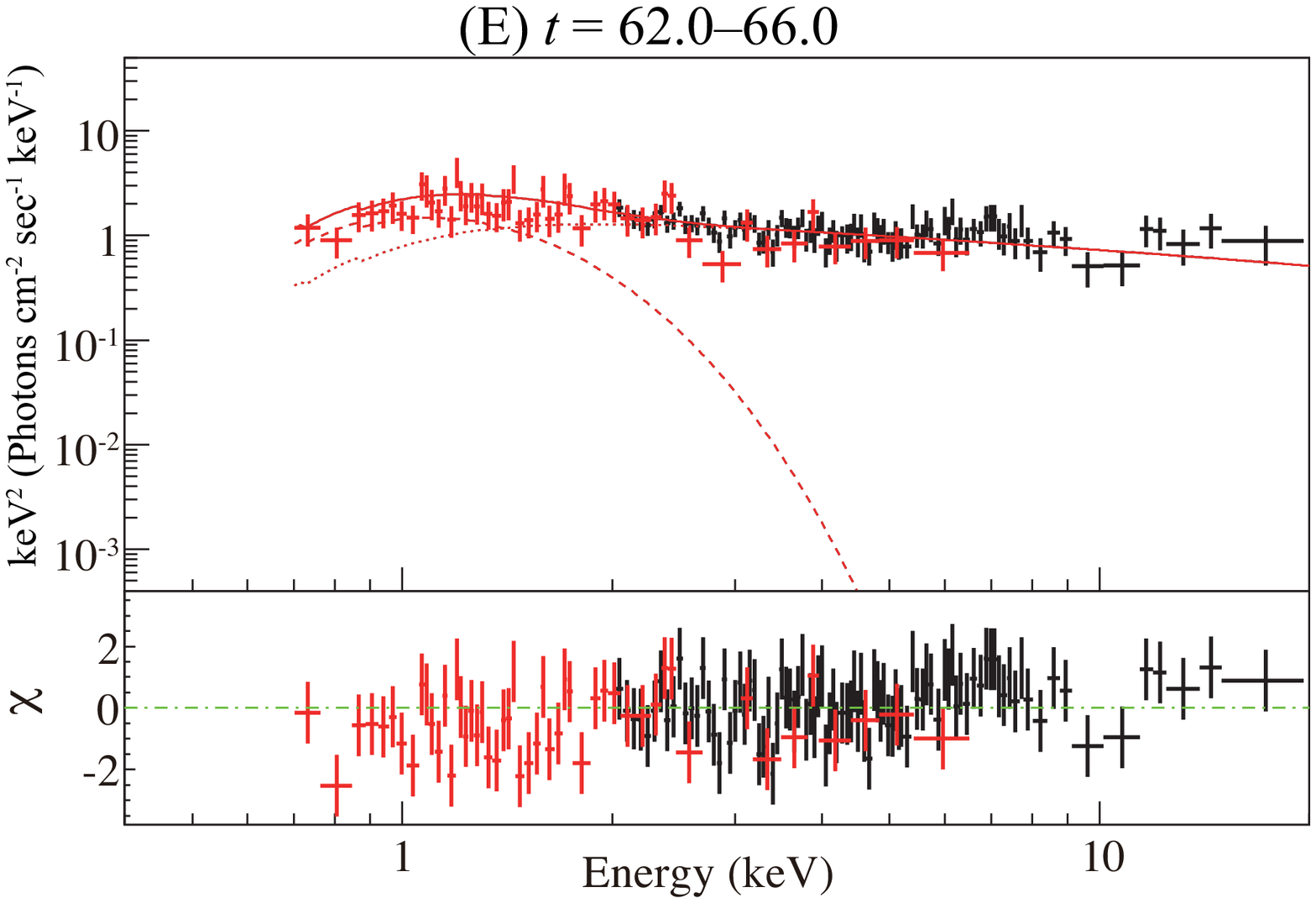}
    \FigureFile(55mm, 55mm){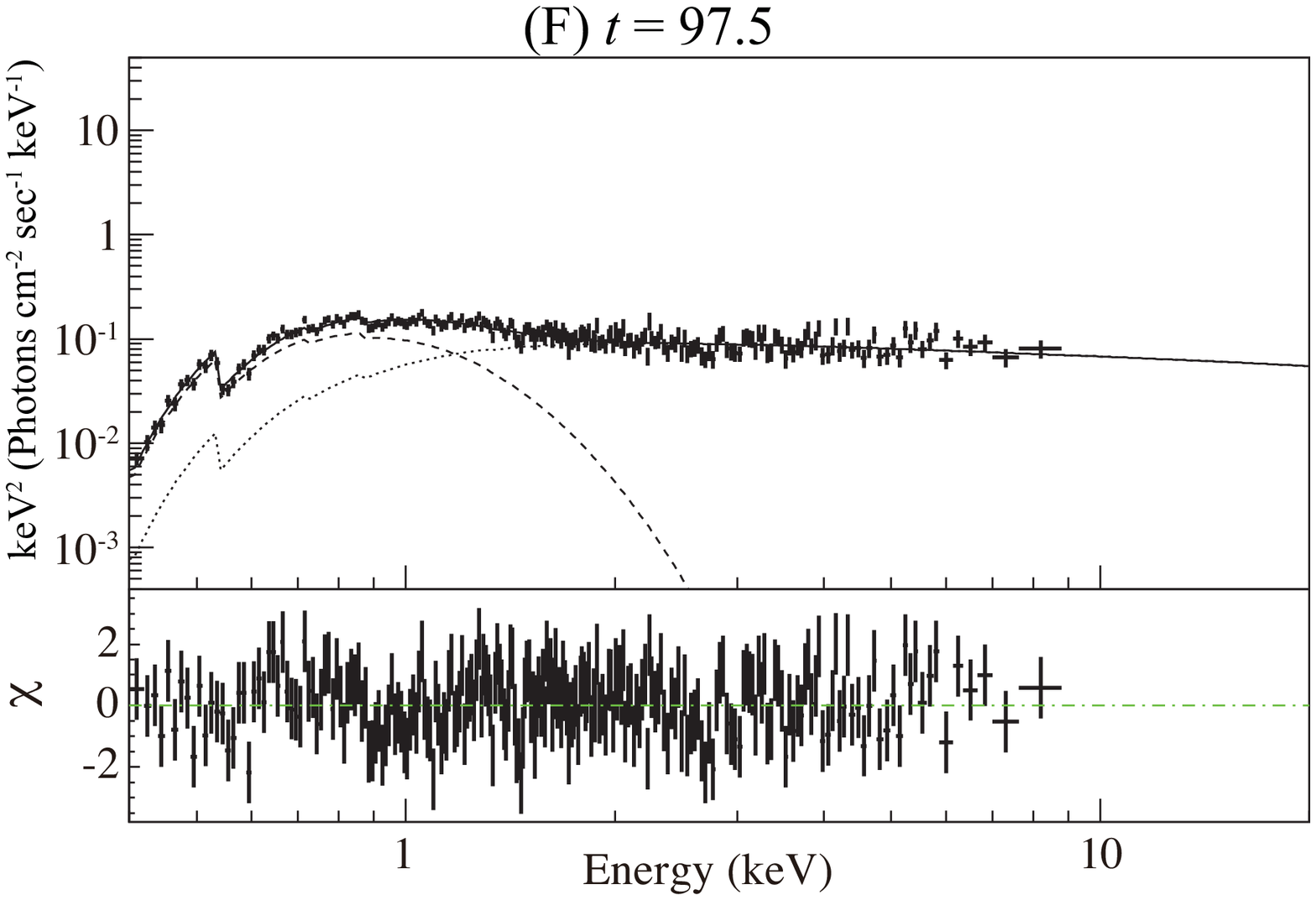}
    \FigureFile(55mm, 55mm){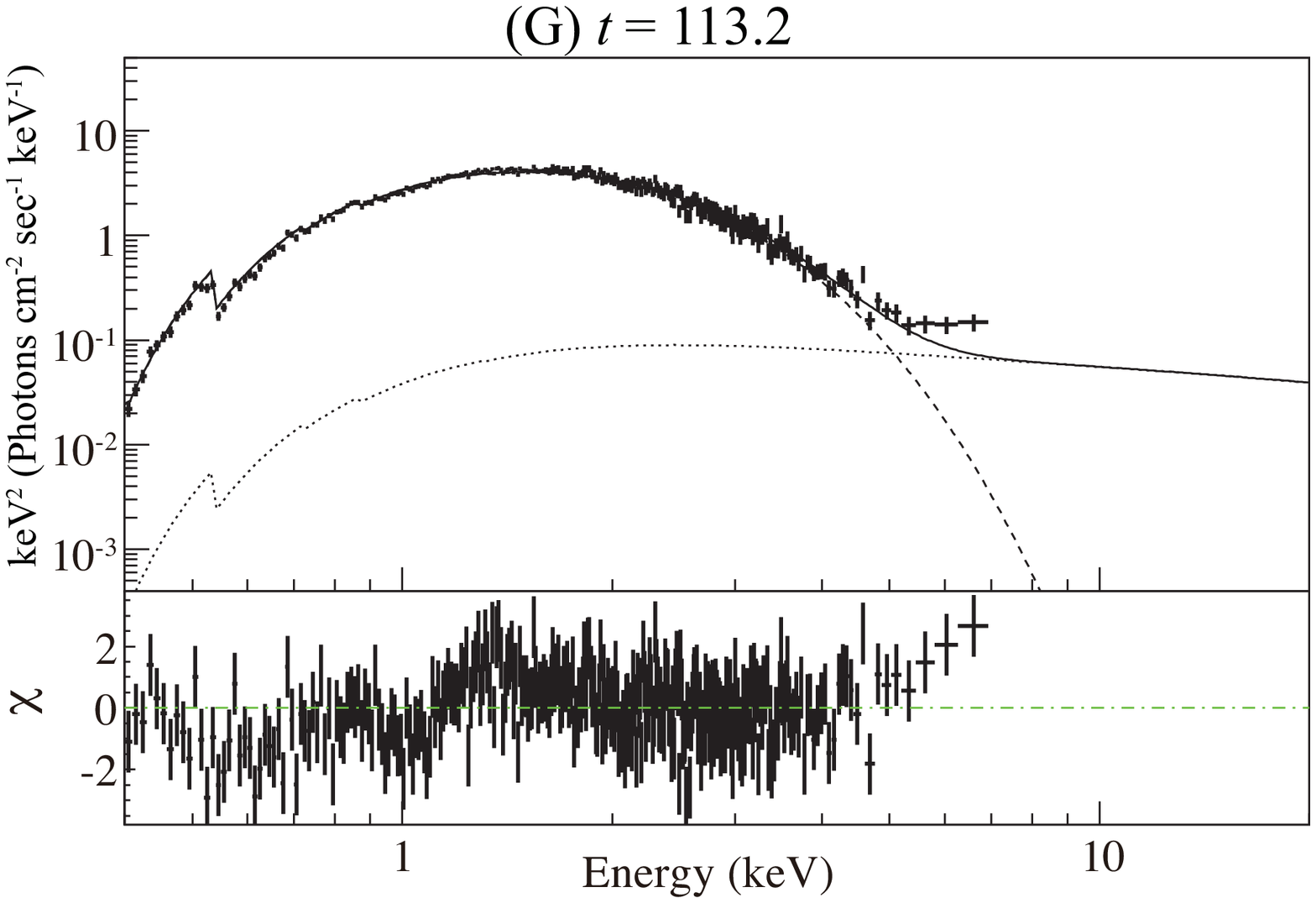}
    \FigureFile(55mm, 55mm){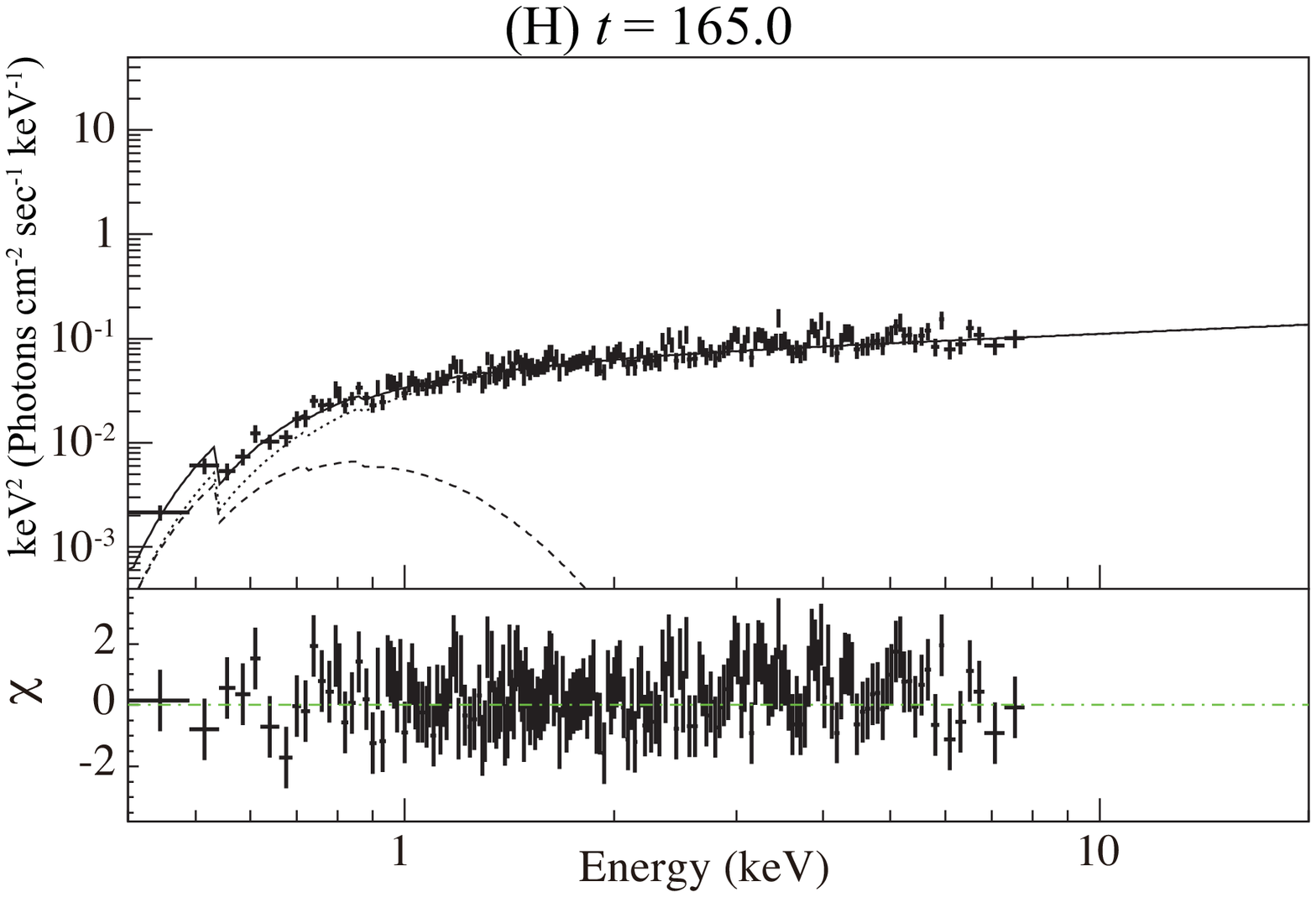}
    \FigureFile(55mm, 55mm){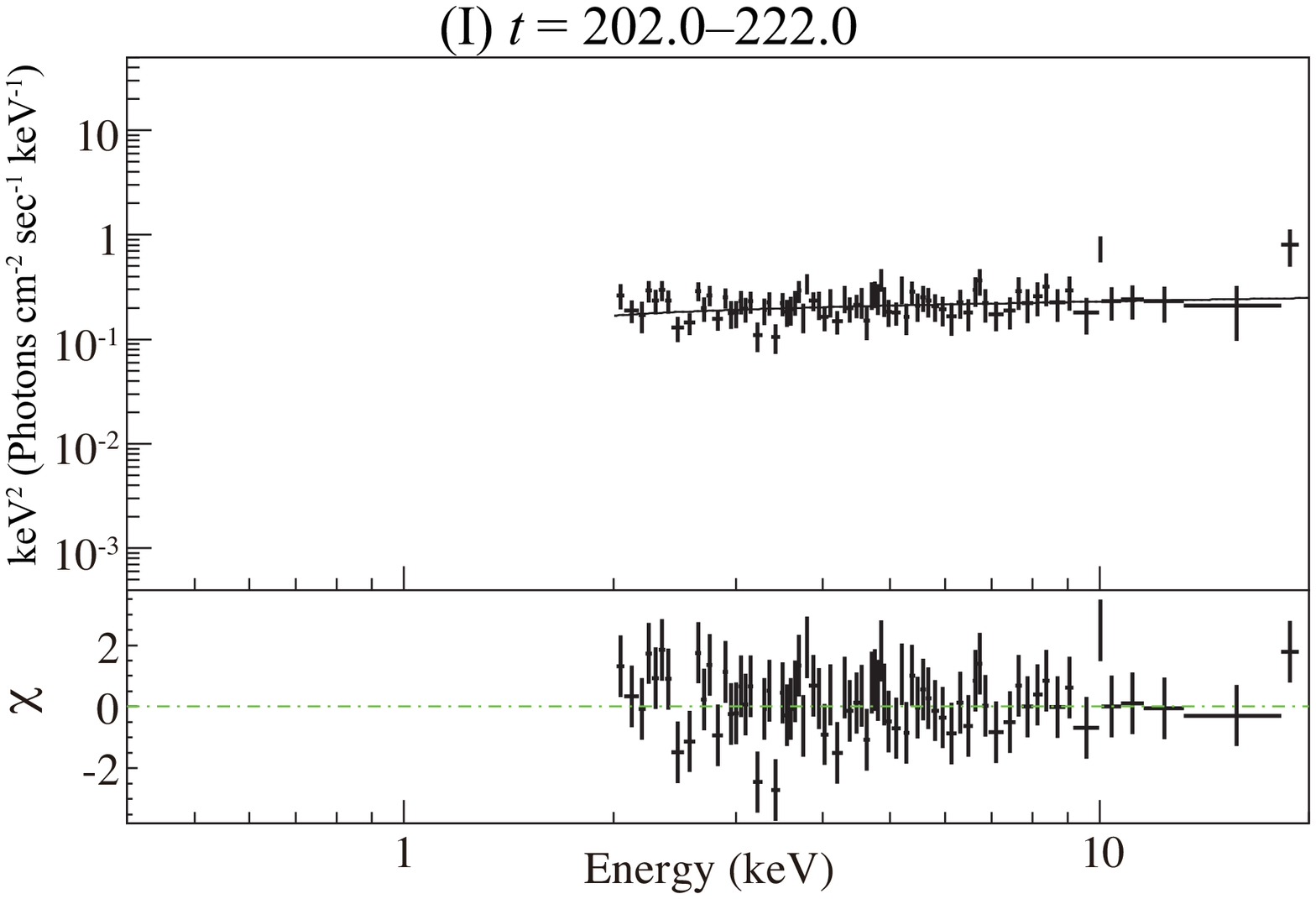}
  \end{center}
  \caption{
{\maxi}/GSC and SSC $\nu$F$\nu$ spectra accumulated in the range of
A) $t$=0.0--1.0 d, 
C) $t$=12.0--13.0 d, 
D) $t$=18.0--19.0 d, 
E) $t$=62.0--66.0 d and 
I) $t$=202.0--222.0 d, 
and {\swift}/XRT snap-shot $\nu$F$\nu$ spectra taken at, 
B) $t$=1.6 d, 
F) $t$=97.5 d, 
G) $t$=113.2 d and 
H) $t$=165.0 d.
}
\label{fig:SPECTRUM}
\end{figure*}

We investigate X-ray spectral evolution during the entire outburst
using {\maxi} data with a continuous time coverage and {\swift}
XRT data with a high sensitivity.  
Because daily {\maxi} data taken by GSC and SSC do not have
enough photon statistics except for the initial bright phase, we
accumulated spectral data for periods of several days for each of GSC
and SSC.  If both GSC and SSC data are available in some time period,
they were fitted simultaneously.
{\swift} XRT spectra were accumulated for each pointing
observations.
Figure \ref{fig:SPECTRUM} presents representative X-ray spectra in
a $\nu F\nu$ form.

Since the compact object is suggested to be a black hole from the
light curve, we employed a model consisting of a Multi-Color
Disk-blackbody (MCD; \cite{mcdmodel}), {\em diskbb} in {\xspec}
terminology, and its Compton up-scattered component with hot thermal
electrons, {\em nthcomp}.
To account for an interstellar absorption, we applied the
Tuebingen-Boulder ISM absorption model ({\em TBabs}) with an abundance
model of \citet{abundwilm}.  The model is described by {\em
TBabs$\times$(diskbb+nthcomp)} with free parameters of the neutral
hydrogen column density ($\nh$), innermost temperature ($\Tin$) and
normalization of the {\em diskbb}, photon index ($\Gamma$), electron
temperature ($kT_{\rm e}$) and a normalization of the {\em nthcomp}.
We fixed $kT_{\rm e}$ at 50 keV because it cannot be constrained in
the observed X-ray band.  Since some observations in the bright phases
indicated edge-like residuals between data and model probably due to
error in the response function, we assigned systematic error of 5\%
into each energy bin corresponds 1.6-2.2 keV of the spectra.

Firstly, we fitted the model to all extracted XRT spectra allowing
$\nh$ free.  The best-fit $\nh$ values were mostly consistent with
$0.41\times 10^{22}$ cm$^{-2}$ during the soft-state periods (SS1 and
SS2).  In the hard state, the value could not be constrained because
the soft X-ray continuum is not well determined.  Hence, we adopted
the best-fit $\nh$ in the soft state, 0.41$\times 10^{22}$ cm$^{-2}$,
as the intrinsic interstellar absorption.
In the SS, $\Gamma$ cannot be determined for each spectrum due
to the limited effective area of GSC, SSC and XRT in the higher energy
band.  Therefore, we fixed $\Gamma$ at 2.4 as typical value in
SS \citep{review}, and we left $\Gamma$ free in the HS.  In some of
the HS spectra observed by GSC alone, the direct MCD component
represented by {\em diskbb} was not significantly detected due to the
poor efficiency in the lower energy band.  In this case, we removed {\em
diskbb} and assigned the typical $\Tin$=0.15 keV for a seed photon of
{\em nthcomp}.  We confirmed that the assumed $\Tin$ 
does not change the best-fit $\Gamma$ in the 2-20 keV band.

\begin{figure}
  \begin{center}
    \FigureFile(90mm,90mm){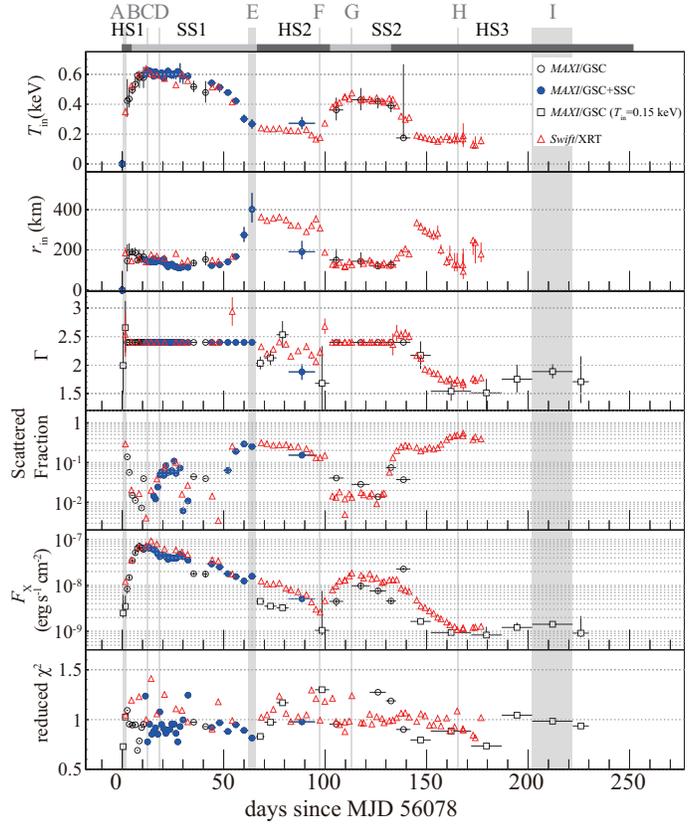}
  \end{center}
  \caption{
  Time evolution of spectral parameters derived from spectral fit.
  Circler points are {\maxi}/GSC (open black) and GSC plus SSC (filled blue) data
  analyzed with {\em TBabs$\times$(diskbb+nthcomp)} model, and triangular points are those with {\swift}/XRT data. 
  Square points show that with GSC data but $\Tin$=0.15 keV is assumed.
  The panels from the top indicate, innermost temperature $\Tin$ (keV), 
  innermost radius $\rin$, photon Index $\Gamma$,
  fraction of {\em nthcomp} component, 
  unabsorbed model flux in 0.01--100 keV $\fx$ and
  reduced $\chi^2$.
  Time periods corresponding to HS1--3, SS1-2 and A--H in figure \ref{fig:SPECTRUM} are shown above.
   }\label{fig:SPECPAR}
\end{figure}

The fits to all the extracted spectra were mostly acceptable.
Table \ref{table:specana} summarizes
the best-fit parameters for the representative spectra  
in figure \ref{fig:SPECPAR}, where the best-fit models are overlaid 
on the data.
Assuming that the Compton up-scattering process is
isotropic and preserves the number of photons, we estimated
innermost radius $\rin$ of the MCD model with a following equation \citep{rineq}
\begin {eqnarray}
\label{rin_nthcomp}
\lefteqn{ F^\mathrm{p}_\mathrm{disk}+F^\mathrm{p}_\mathrm{thc} 2 \cos i} \nonumber \\
\ & = 0.0165 \biggl[ \frac{r^{2}_\mathrm{in} \cos i}{(D/10\ \mathrm{kpc})^{2}} \biggr] \biggl( \frac{T_\mathrm{in}}{1\ \mathrm{keV}} \biggr)^3 \mathrm{photons}\  \mathrm{s}^{-1} \mathrm{cm}^{-2} ,
\end {eqnarray}
where the $F^\mathrm{p}_\mathrm{disk}$ and $F^\mathrm{p}_\mathrm{thc}$ represent the photon fluxes
from the MCD and Comptonized component, respectively.
We also calculated the fraction of the 
Compton-scattered photon,
$f_\mathrm{sc}$~=~$\frac{F^\mathrm{p}_\mathrm{thc}}{F^\mathrm{p}_\mathrm{disk}+F^\mathrm{p}_\mathrm{thc}}$.
These values are listed in \ref{table:specana}.
Figure \ref{fig:SPECPAR} shows the variations of the spectral
parameters as a function of time.

\begin{table*}
\small
  \caption{Best-fit model parameters obtained from {\maxi}/GSC, SSC and {\swift}/XRT spectra.}
  \label{table:specana} 
  \begin{center}         
  \begin{tabular}{ccccccccc}
  \hline\hline
    Period  & Instruments & $t^*$  &$\Tin$  &$\rin$\footnotemark[$\dagger$] & $\Gamma$ &   $f_\mathrm{sc}$\footnotemark[$\ddagger$] & $\fx$\footnotemark[$\S$] & $\chi^{2}$~/~d.o.f\\
      &   & (day)  & (keV) &(km)  & & &  \\
   \hline
  A & {\maxi}/GSC & 0.0--1.0 & -- & -- & 1.99$_{-0.51}^{+0.59}$ & -- & 5.0$_{-1.3}^{+16.7}$  &  4.4 / 6 \\
  B & Swift/XRT & 1.6 & 0.35$\pm{0.03}$ & 183$_{-19}^{+29}$ & 2.54$_{-0.39}^{+0.38}$ & 0.30  & 12.0$_{-0.3}^{+0.2}$  &  214.9 / 201 \\
  C & {\maxi}/GSC+SSC & 12.0--13.0 & 0.63$\pm{0.01}$ & 141.6$_{-4.2}^{+4.6}$ & 2.4 (fixed) & 0.00  & 69.1$\pm2.5$  &  168.5 / 217 \\
  D & {\maxi}/GSC+SSC & 18.0--19.0 & 0.59$\pm{0.02}$ & 146.4$_{-5.7}^{+5.9}$ & 2.4 (fixed) & 0.05  & 66.4$_{-3.5}^{+3.7}$  &  272.8 / 254 \\
  E & {\maxi}/GSC+SSC & 62.0--66.0 & 0.27$\pm{0.03}$ & 400$_{-64}^{+83}$ & 2.4 (fixed) & 0.25  & 48.3$\pm1.8$  &  116.2 / 143 \\
  F & Swift/XRT & 97.5 & 0.18$\pm{0.01}$ & 303$_{-14}^{+16}$ & 2.26$\pm{0.08}$ & 0.14  & 2.6$\pm0.1$  &  364.3 / 362 \\
  G & Swift/XRT & 113.2 & 0.47$\pm{0.00}$ & 125.8$\pm{1.4}$ & 2.4 (fixed) & 0.01  & 18.5$\pm0.1$  &  480.3 / 388 \\
  H & Swift/XRT & 165.0 & 0.17$\pm{0.04}$ & 128$_{-35}^{+57}$ & 1.73$_{-0.07}^{+0.06}$ & 0.49  & 1.1$\pm0.1$  &  386.6 / 356 \\
  I & {\maxi}/GSC & 202.0--222.0 & -- & -- & 1.88$_{-0.12}^{+0.13}$ & -- & 2.5$_{-0.2}^{+0.3}$  &  72.7 / 74 \\
   \hline
  \end{tabular}
  \end{center}
  \footnotemark[$*$] {Elapsed days since MJD=56078.0}\\
  \footnotemark[$\dagger$]{D=10.0 kpc and i = 0◦ are assumed.}\\
  \footnotemark[$\ddagger$]{Scattering fraction of the model, calculated from photon numbers of {\em diskbb} ($\fdisk^p$) and {\em nthcomp}($F_{\rm thc}^p$) to be $\frac{F_{\rm thc}^p} {F_{\rm thc}^p+\fdisk^p}$. }\\
  \footnotemark [$\S$] {Absorption-corrected model flux in units of $10^{-9} \ergcms$ in 0.01--100~keV band.}
\end{table*}

\subsection{X-ray hardness-intensity diagram}\label{xray_hid}

\begin{figure}
  \begin{center}
    \FigureFile(90mm,90mm){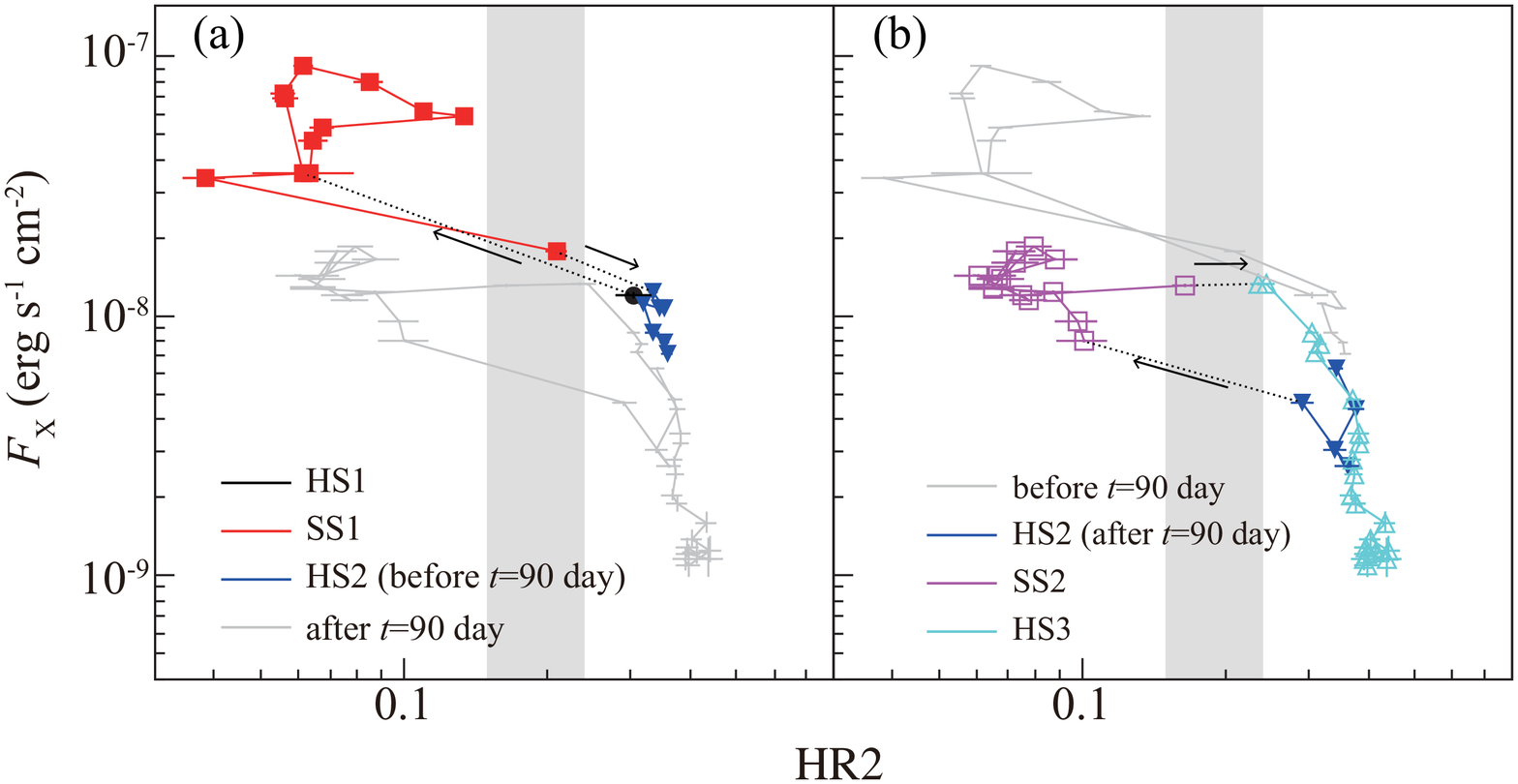}
  \end{center}
  \caption{
  Hardness Ratio (HR2) vs 0.01-100 keV $\fx$ diagram, in the periods 
  (a) from $t$=0 to 90 d and 
  (b) from 90 to the end.
  Data corresponding to the periods SS1, SS2, HS1, HS2 and HS3 are shown 
  in red, magenta, black, blue and cyan, respectively.
  The data points shown in gray color are identical with that of opposite sides.
  We assumed that soft-to-hard transition occurred at HR 
  ranging from 0.15 to 0.25 (shadowed region).
   }
\label{fig:SwiftHID}
\end{figure}

In order to clarify the spectral evolution, hardness-ratio versus flux 
diagrams were plotted in figure \ref{fig:SwiftHID}.
The horizontal axis and vertical axis represent {\swift}/XRT 3-5/5-10 keV HR and $\fx$ 
obtained with model fits, respectively.
The figure is split into 
(a) HS1, SS2 and before $t$=90 d of HS2, 
(b) after $t$=90 d of HS2, SS2 and HS3.
Comparing the diagram with those obtained with past observations
of black holes (e.g. \cite{qcurve}, \cite{nkhr1752a}), 
we found two unusual features. 
One is that it exhibited  hard-to-soft and soft-to-hard spectral state 
transitions twice, as seen in figures \ref{fig:SwiftHID}(a) and (b).
The other is transition luminosities. Usually Hard-to-soft transition occurs 
at higher luminosity than soft-to-hard transition. 
However, in the case of SS2 to HS3 transition occurred at 
$\sim$twice brighter luminosity than that of HS2 to SS2 transition.
Nevertheless, we found that SS1 to HS2 and SS2 to HS3 transitions occur at a similar X-ray flux.
If we assume that the transitions occur at HR of 0.15--0.25, the corresponding flux becomes 1.28--2.33 $\ergcmse$.

\subsection{X-ray and UV correlation}\label{x-uv_corr}

\begin{figure}
  \begin{center}
    \FigureFile(90mm,90mm){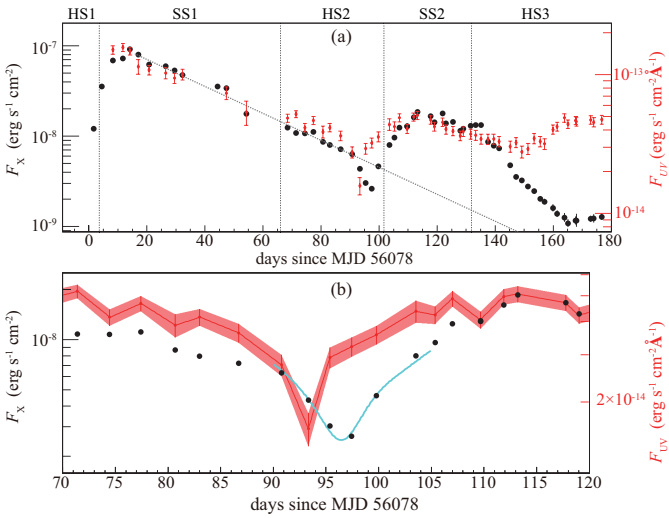}
  \end{center}
  \caption{
  (a) X-ray and UV light curves until $t=$180 d, 
  black and red plots indicate {\swift}/XRT 0.01--100 keV $\fx$ derived spectral fits
  and {\swift}/UVOT UVM2 flux respectively. 
  The fluxes are corrected for interstellar absorption and extinction 
  with parameters shown in section \ref{xray_spec}, \ref{uv_spec}.
  (b) Same as above data but zoomed in $t$=70--120 d, and filled region shows the logarithmic 
  interpolated UVOT data with its statistical error.
  The best-fit model (see text) is shown in cyan line.}
\label{fig:CCF}
\end{figure}
\begin{figure}
  \begin{center}
    \FigureFile(90mm,90mm){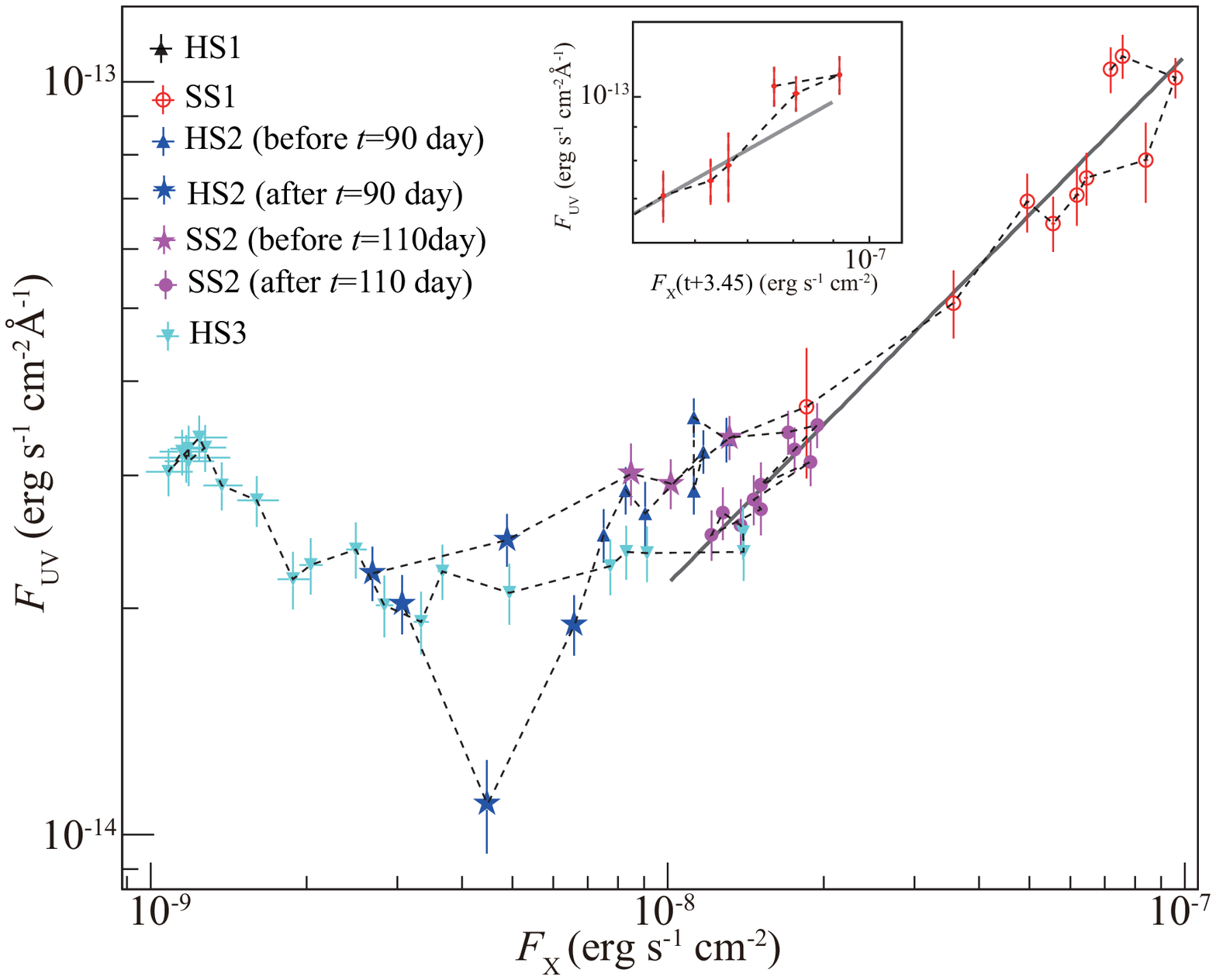}
  \end{center}
  \caption{Scatter plot of $\fx$ versus $\fuv$, 
  with the same color-indexing as figure \ref{fig:SwiftHID}.
  The data points were connected with a dashed line in the order of time-series.
  The solid line indicates $\fuv \propto \fx^{0.70}$ obtained for SS1 and SS2.
  Sub figure is modified for a UV--X-ray delay
  interpolating $\fx$, solid line is the same as main plot.
  }
\label{fig:FLUX_UVM2}
\end{figure}

In this section, we investigate relation between X-ray and UV emission.
Since optical/UV emission is reddened by the dust lying in the line of sight toward the source, 
we estimated the extinction factor utilizing empirical relation given by \citet{guver_extinction} as 
$\nh = (2.21\pm0.09) {\rm A_v}\times10^{-21}$ cm$^{-2}$.
From $\nh = 0.41 \times 10^{22}$ cm$^{-2}$ obtained by X-ray spectrum, and assuming 
${\rm A_v} = 3.1 E(B-V)$ \citep{extinction_savage}, we obtained $E(B-V) = 0.598$. 

We have overlaid X-ray and UV light curves in figure \ref{fig:CCF}(a). 
$\fx$ is identical to that indicated in figure \ref{fig:SPECPAR},
and $\fuv$ is the same data; but we converted UVM2 Magnitude into extinction-corrected flux.
We found that $\fx$ and $\fuv$ correlated nicely during exponentially decay phases, 
that is $t=$ 10--90 d and $t=$ 110--130 d.
The e-folding time of $\fx$ and $\fuv$ flux was 30 and 56 days for $t$=10--90 d, 
and 37 and 51 days for $t$=110-130 d, respectively. 
We obtained e-folding time of $\fuv$ to be about twice larger than that of $\fx$.
However, we can find apparent mismatches on $t=$ 90--110 d, and after $t$=140 d.
We show scatter plots between $\fuv$ and $\fx$ in figure \ref{fig:FLUX_UVM2}.
Excluding the period $t=$ 90--110 d (shown with star points), the relation is 
approximated by $\fuv \propto \fx^{0.70\pm0.02}$ in the SS1 and SS.

Figure \ref{fig:CCF}(b) zooms up the light curves around $t=$ 90--110 d. 
We can find the dip with a time difference
between $\fx$ and $\fuv$ and subsequent $\fuv$ excess.  In order to
estimate the time difference from $\fuv$ to $\fx$ on the dip, we
perform a model fit.  The model is described by a logarithmically
interpolated $\fuv$ light curve (red region in
figure \ref{fig:CCF}(b)) which is convolved with a Gaussian kernel.
Using this model, we fitted $\fx$ light curve from $t$=90 to 110 d
assuming $\fuv \propto \fx^{0.45}$ obtained from the corresponding
period of figure \ref{fig:FLUX_UVM2}.  Eventually, we obtained a
profile of time delay from $\fuv$ to $\fx$ to be centered at
3.5$\pm0.2$ days and a full-width at half-maximum (FWHM) of
3.8$_{-0.7}^{+0.8}$ days.  We indicated a best-fit model with a
cyan line in \ref{fig:CCF}(b).

\subsection{Joint UV and X-ray spectral analysis}\label{uv_spec}

In order to investigate origin of the UV emission, we performed {\swift}/XRT spectral 
analyses including {\swift}/UVOT data.
We first examined the SS spectra taken at $t$=17.0 d with six UVOT filters.
We used a {\em redden$\times$TBabs$\times$(diskbb+thcomp)} model in {\xspec} 
where the {\em redden} model represents the extinction.
We found that the UV/optical flux is 2.5-4 times larger than that extrapolated from the MCD. 
The fitting is far from acceptable level at 3.54 ($\chi^2$/d.o.f.=1140.1/322).

Next we employed the {\em diskir} model \citep{J1817_G2008} which calculates a reprocessed energy spectrum 
from an irradiated disk by X-rays from the {\rm diskbb} and {\rm nthcomp} model components.
We used a {\em redden$\times$TBabs$\times$diskir} model
where we fixed $E(B-V)$ at 0.598, the fraction of the thermalized component
$f_\mathrm{in}$ at 0.1, and a radius of the illuminated disk 
$r_\mathrm{irr}$ at 1.1 $\rin$ (e.g. \cite{J1817_G2009}, \cite{1305shidatsu}).

\begin{figure*}
  \begin{center}
    \FigureFile(50mm,50mm){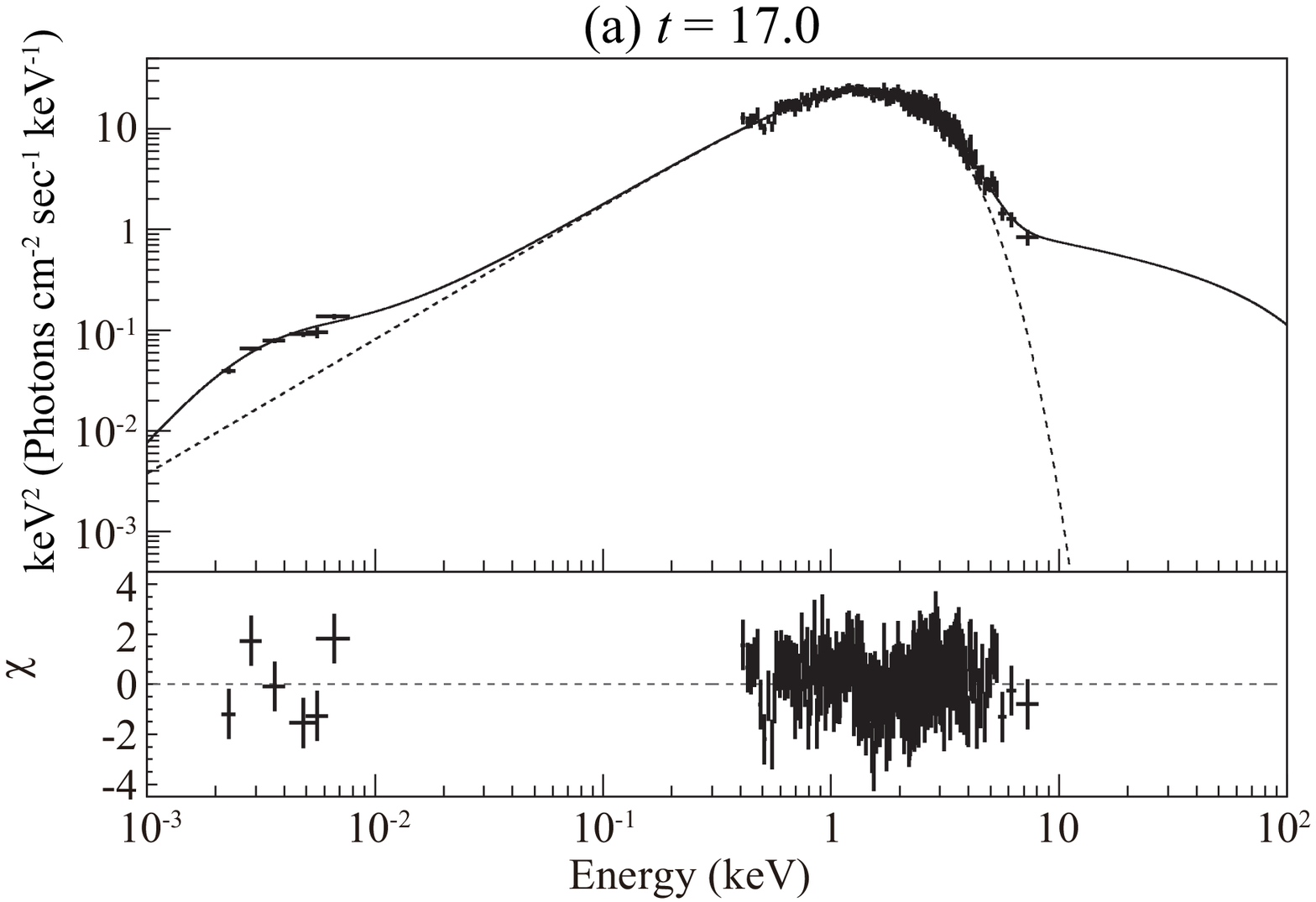} 
    \FigureFile(50mm,50mm){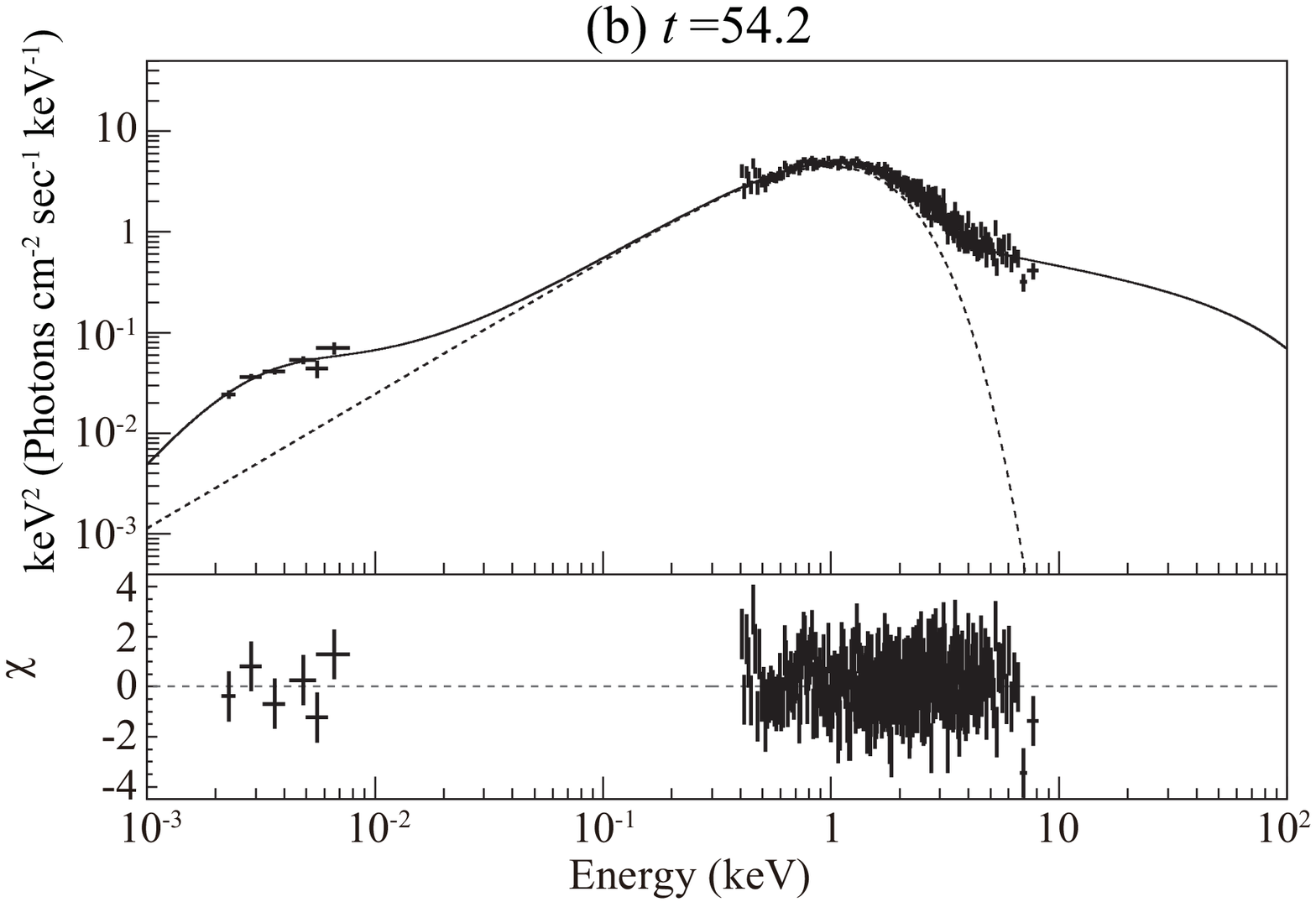} 
    \FigureFile(50mm,50mm){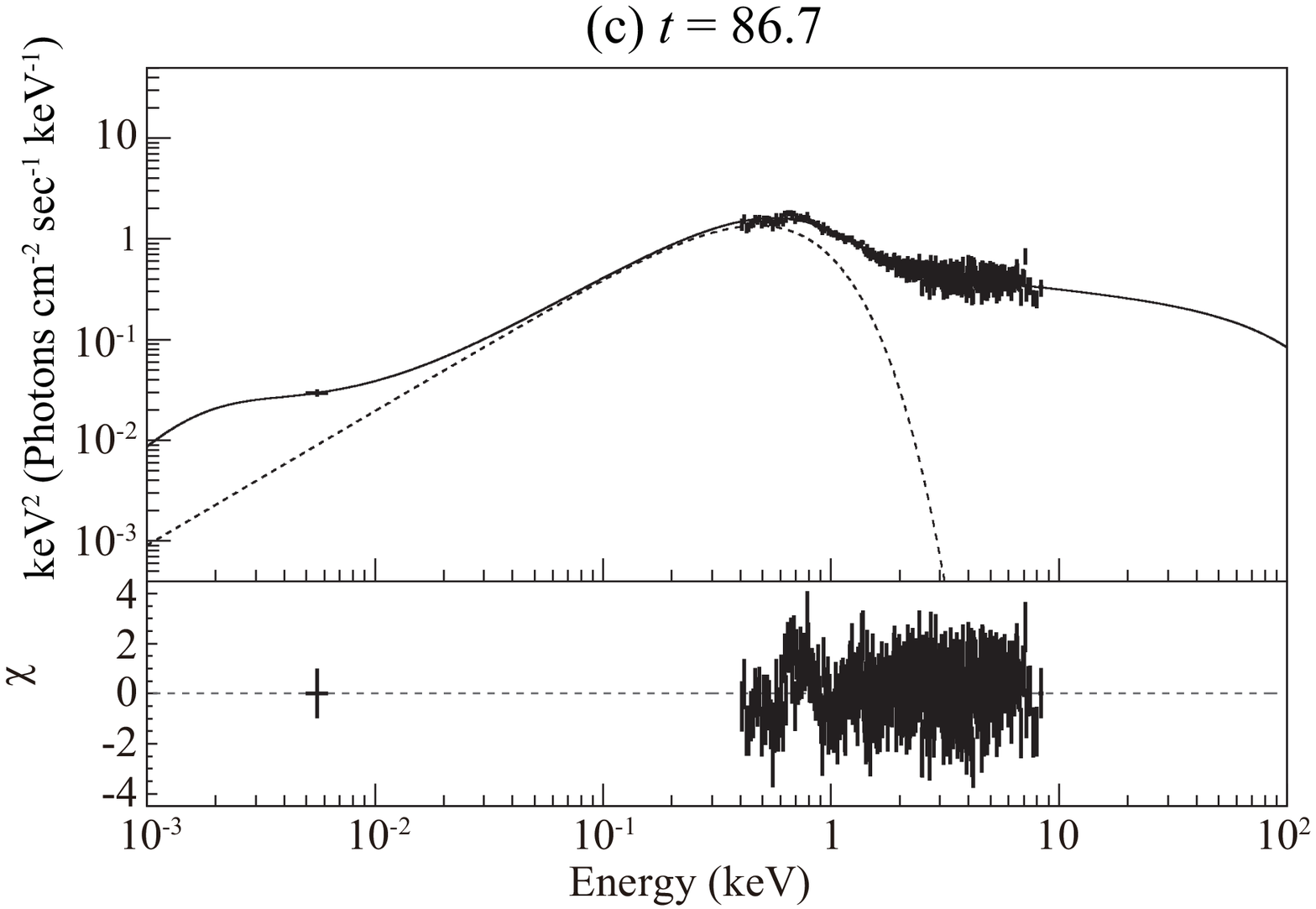} 
    \FigureFile(50mm,50mm){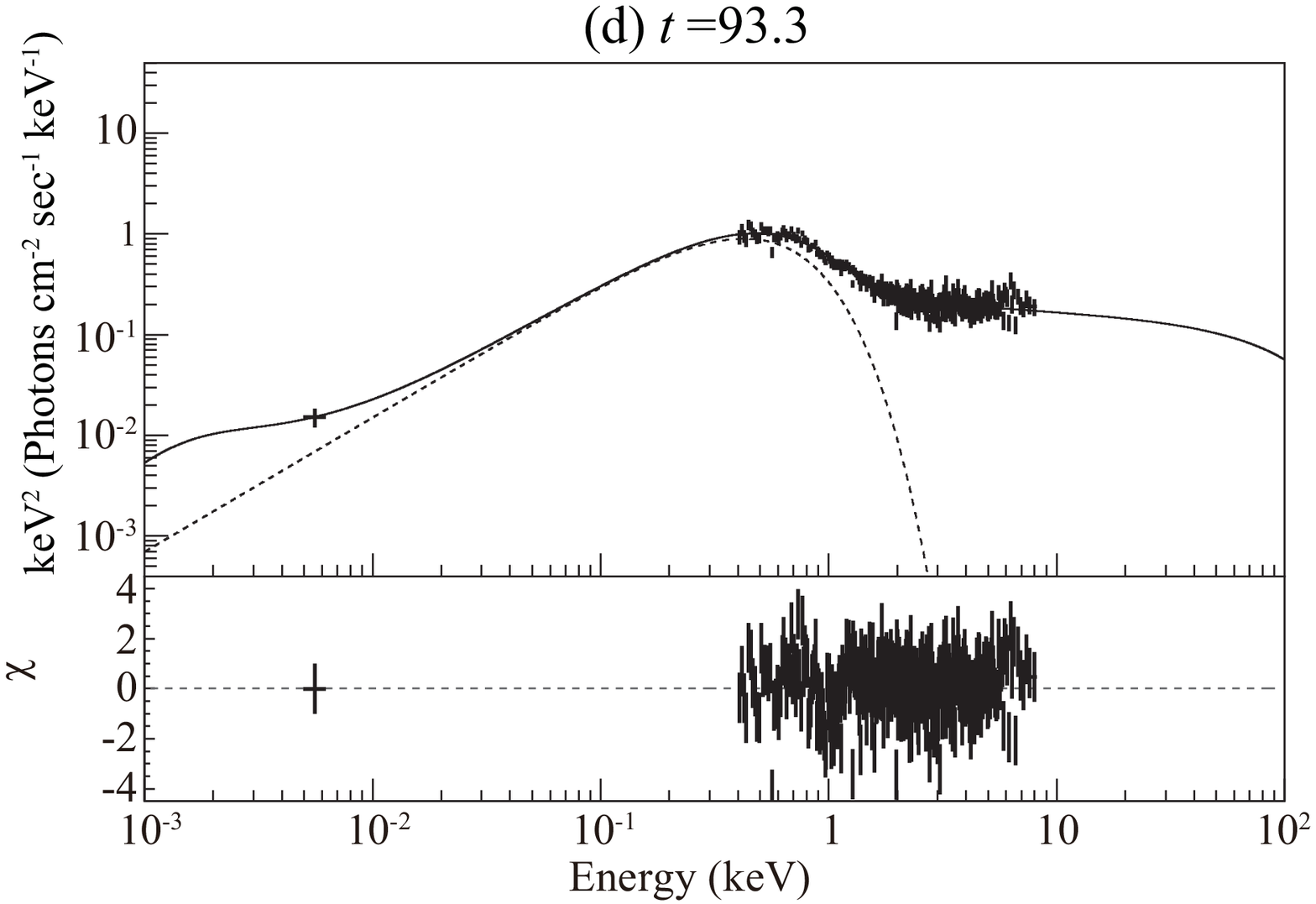} 
    \FigureFile(50mm,50mm){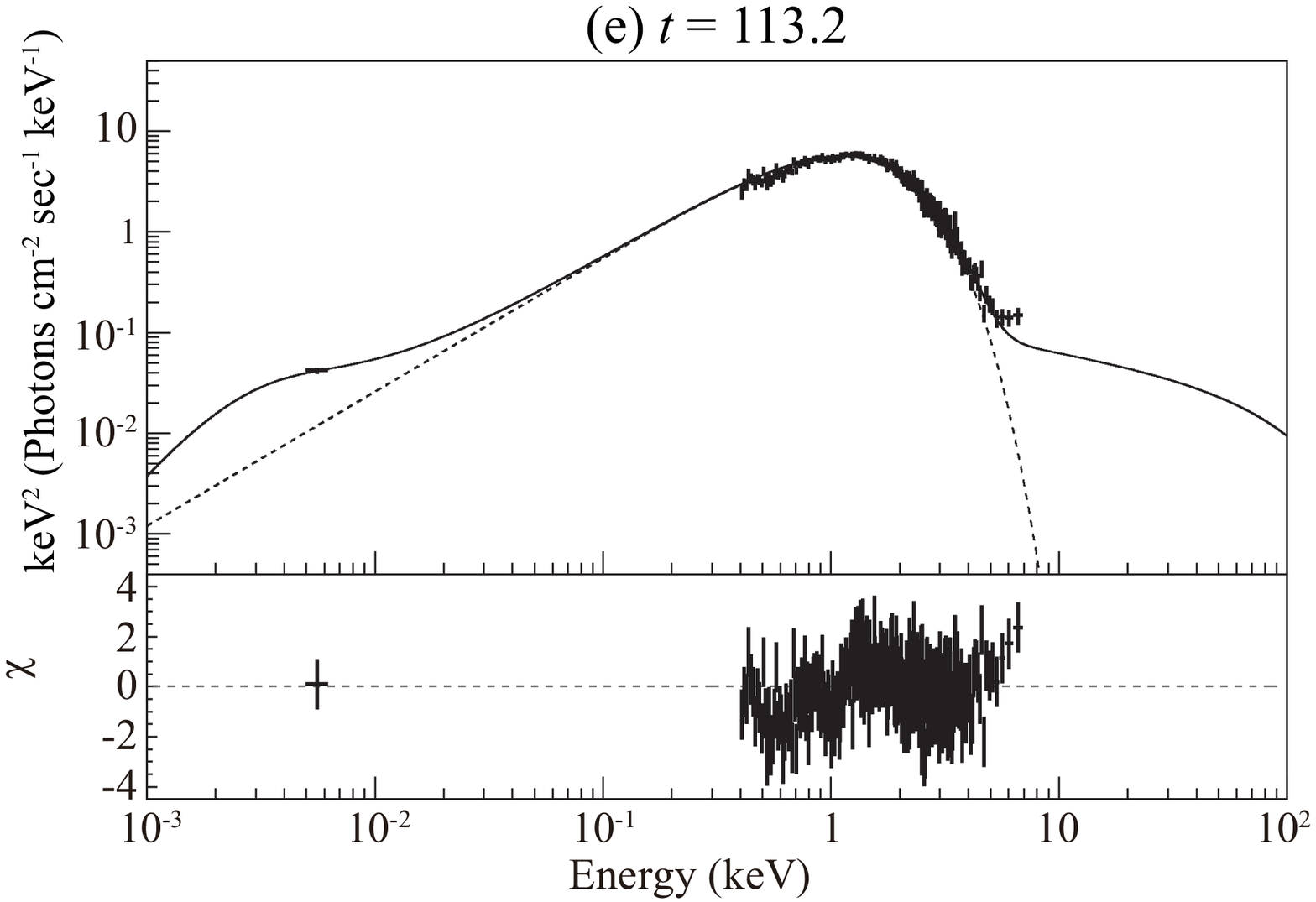} 
    \FigureFile(50mm,50mm){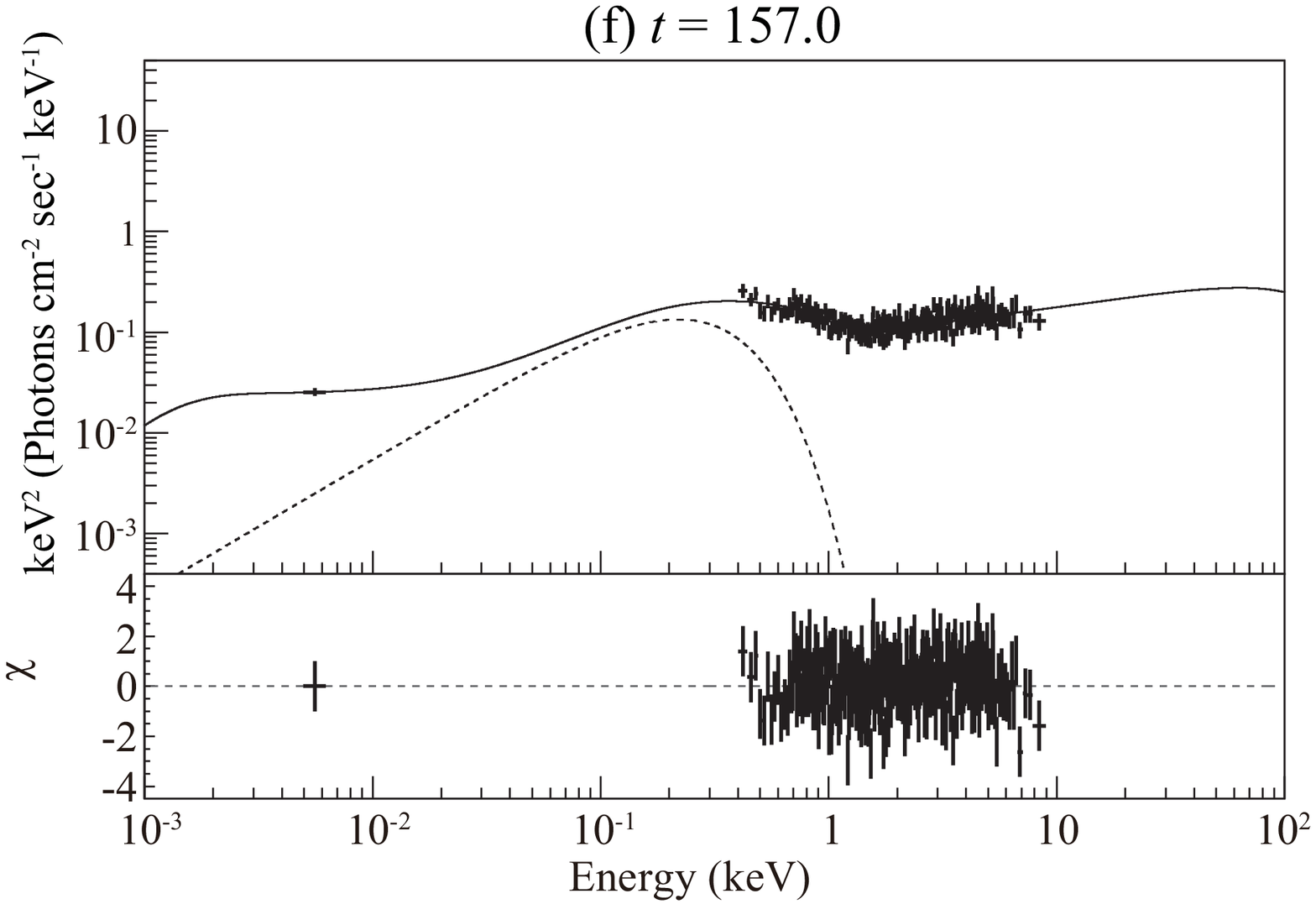} 
  \end{center}
  \caption{
   UV and X-ray $\nu F\nu$ spectra obtained by {\swift} UVOT and XRT, respectively, 
   at $t=$ 17.0 d, 54.2 d, 86.7 d, 86.7 d, 93.3 d, 113.2 d, and 157.0 d.
   Each upper panel shows observed and model spectra corrected for interstellar absorption and extinction. 
   The solid line indicates a total reprocessed disk spectrum, and the dashed line shows an intrinsic disk spectrum.
   Bottom panels represent residuals between the UV and X-ray data and the reprocessed model. 
  }
\label{fig:uvot_sed}
\end{figure*}

The fit was acceptable yielding $\chi^2$/d.o.f. = 344.2/320.
Unabsorbed and de-reddened $\nu F\nu$ spectra of the irradiated disk
model and the data are shown in figure \ref{fig:uvot_sed}(a).
The model has free parameters of $\Tin$, $\rin$, an outer disk radius
$r_\mathrm{out}$, a luminosity ratio of the Compton tail to the unilluminated (inner) disk
$L_\mathrm{c}$/$L_\mathrm{d}$, and a fraction of bolometric flux thermalized in the outer disk
 $f_\mathrm{out}$.  The parameters in the X-ray band
were almost consistent with those obtained with the {\em diskbb} model, 
$\Tin$=0.59$\pm0.01$ keV, $\rin$=170.0$_{-4.5}^{+4.6}$ km, 
$\log_{10}$($r_\mathrm{out}$/$r_\mathrm{in}$)=$4.32_{-0.07}^{+0.08}$,
and $f_\mathrm{out}$=$0.90_{-0.11}^{+0.13} \times 10^{-3}$.

\begin{figure}
  \begin{center}
    \FigureFile(80mm,80mm){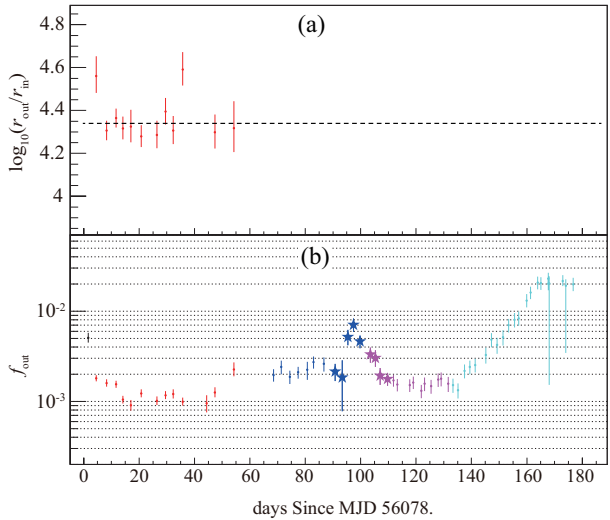}
  \end{center}
  \caption{
      Change of (a) $r_\mathrm{ out}$ and (b) $f_\mathrm{ out}$  as a function of time 
      obtained with reprocessed disk model.
      The plot is created in the same manner as figure \ref{fig:FLUX_UVM2}.
      }
\label{fig:routfout}
\end{figure}

Assuming that the UV-excess fluxes are entirely produced by the
disk irradiation, we applied this model for all the {\swift} observations.
Since the spectrum with a single UVOT band data cannot give constraints 
on $r_\mathrm{out}$ and $f_\mathrm{ out}$ at the same time.
Thus we first analyzed spectra which have 6 UVOT bands allowing 
both $r_\mathrm{out}$ and $f_\mathrm{out}$ to be free.
As shown in figure \ref{fig:routfout}(a), $\log_{10}$ ($r_\mathrm{out}/\rin$) are found to 
be constant at $\sim$4.33.
Then we fixed $r_\mathrm{out}$ at this value, and fit the other spectra.
The results of $f_\mathrm{out}$ are shown as a function of time (figure \ref{fig:routfout} b).
The representative spectra and these best-fit parameters are summarized in 
figures \ref{fig:uvot_sed} and table\ref{table:uvspecana}.

\begin{table*}
  \caption{Best-fit model parameters obtained from a reprocessed model with {\swift}/XRT and UVOT spectra.}
  \label{table:uvspecana} 
  \begin{center}         
  \begin{tabular}{cccccccc| c}
  \hline\hline
    Time$^{*}$ & $\Tin$  & $\rin$$^{\dagger}$ & $\Gamma$ &  $F_{\rm c}/F_{\rm d}$$^{\ddagger}$ & $f_{\rm out}$$^{\S}$ 
      & $\log_{10} (r_{\rm out}/\rin$) $^{\|}$ & $\chi^{2}$~/~d.o.f. &  $f_{\rm UV}^{\rm obs}$ / $f_{\rm UV}^{\rm MCD}$$^{\#}$ \\
        & (keV) & (km) &  &  & $\times$ 10$^{-3}$ &  &  &  \\
   \hline
      17.0 & 0.59$\pm{0.01}$ & 170.0$_{-4.5}^{+4.6}$ & 2.40 (fixed) & 0.06$\pm{0.01}$ & 0.90$_{-0.11}^{+0.13}$ & 4.32$_{-0.07}^{+0.08}$ & 343.8 / 320 &   $2.56 \pm 0.31$\\ 
      54.2 & 0.42$\pm{0.01}$ & 148.0$_{-4.6}^{+4.8}$ & 2.40 (fixed) & 0.23$\pm{0.01}$ & 2.25$_{-0.35}^{+0.46}$ & 4.32$_{-0.11}^{+0.13}$ & 340.4 / 336 & $3.86 \pm 0.77$ \\ 
      86.7 & 0.20$\pm{0.01}$ & 349$_{-19}^{+20}$ & 2.22$\pm{0.05}$ & 0.64$\pm{0.02}$ & 2.60$\pm{+0.47}$ & 4.33 (fixed) & 444.9 / 431 & $3.52 \pm 0.29$ \\ 
      93.3 & 0.18$\pm{0.01}$ & 354$_{-24}^{+26}$ & 2.14$\pm{0.07}$ & 0.51$\pm{0.03}$ & 1.84$_{-1.06}^{+1.01}$ & 4.33 (fixed) & 411.8 / 323 & $2.31 \pm 0.48$ \\ 
      113.2 & 0.47$\pm{0.01}$ & 128.2$\pm{2.2}$ & 2.40 (fixed) & 0.02$\pm{0.01}$ & 1.53$\pm{0.24}$ & 4.33 (fixed) & 388.1 / 324 &  $3.55 \pm 0.25$  \\ 
      157.0 & 0.09$\pm{0.01}$ & 517$_{-59}^{+71}$ & 1.72$_\pm{0.05}$ & 4.04$_{-1.05}^{+1.88}$ & 8.35$_{-1.47}^{+1.60}$ & 4.33 (fixed) & 252.2 / 248 &  $17.76 \pm 1.58$ \\ 
   \hline
  \end{tabular}
  \end{center}
  \par\noindent
  \footnotemark[$*$] {Elapsed days since MJD=56078.0}
  \par\noindent
  \footnotemark[$\dagger$]{D=10.0 kpc and $i$ = 0$^\circ$ (face-on) are assumed.}
  \par\noindent
  \footnotemark[$\ddagger$] ratio of luminosity in the Compton tail to that of the unilluminated disk
  \par\noindent
  \footnotemark [$\S$] fraction of bolometric flux which is thermalized in the outer disk
  \par\noindent
  \footnotemark [$\|$] $\log_{10}$ of the outer disk radius in terms of the inner disk radius
  \par\noindent
  \footnotemark [$\#$] ratio of observed UVM2 band flux to that calculated from MCD
  \par\noindent
  \footnotemark [$\**$] fraction of luminosity in the Compton tail which is thermalized in the inner disk is fixed at 0.1
  \par\noindent
  \footnotemark [$\dagger\dagger$] radius of the Compton illuminated disk in terms of the inner disk radius is fixed at 1.1
\end{table*}

\section{Discussions}

\subsection{Summary of the results}

We monitored the entire outburst from MAXI~J1910--057 with {\maxi} and {\swift}, 
as the light curves in figure \ref{fig:LC}.
As the spectrum changed dramatically several times, we separated 
the periods into the HS1, 2, 3, SS1 and SS2.
As described in section \ref{xray_spec}, we analyzed all {\t MAXI} stacked spectra and {\swift}/XRT 
with a MCD plus a thermal Comptonization model.

MAXI~J1910--057 broke out on MJD 56078 ($t=0$ d).
On the first day (HS1), it showed a hard X-ray spectrum, and 
then it rapidly changed to a soft spectrum (SS1).
The X-ray flux reached a maximum at $t=10$ d.
The hard X-ray increased due to an enhancement of Comptonization around $t=20$ d.
We found a secondary peak around $t=50$ d
in 2--6 keV {\maxi}/GSC light curve, which is a factor of 2 higher than 
that extrapolated from $t=10-40$ d.
However, we consider that it is an induced structure in this energy band 
due to the hard-to-soft spectral change.
Actually, the 0.01--100 keV bolometric flux between $t=10$ d and 90 d showed 
a rather steady exponential decay with a time constant of $\sim$30 days. 
We note that $\Gamma$ remained above 2.0, which is higher than that 
compared with a typical value in the HS of 1.4-1.7 \citep{review2}. 
Hence, the HS2 could be characterized as so-called ``hard intermediate state''. 

A dip structure was found around $t=95$ d in X-ray flux; it is followed by 
a bright re-flare (increased by a factor of $\sim$ten from the dip) peaking at $t=113$ day.
The X-ray spectrum was confirmed to be the soft state (SS3), meanwhile.
Remaining the hard spectrum (HS3), the outburst lasted with a flat tail until 
$t\sim230$ d, and then the X-ray flux declined steeply .

\subsection{Structure and evolution of the inner accretion disk}

\begin{figure}
  \begin{center}
    \FigureFile(90mm,90mm){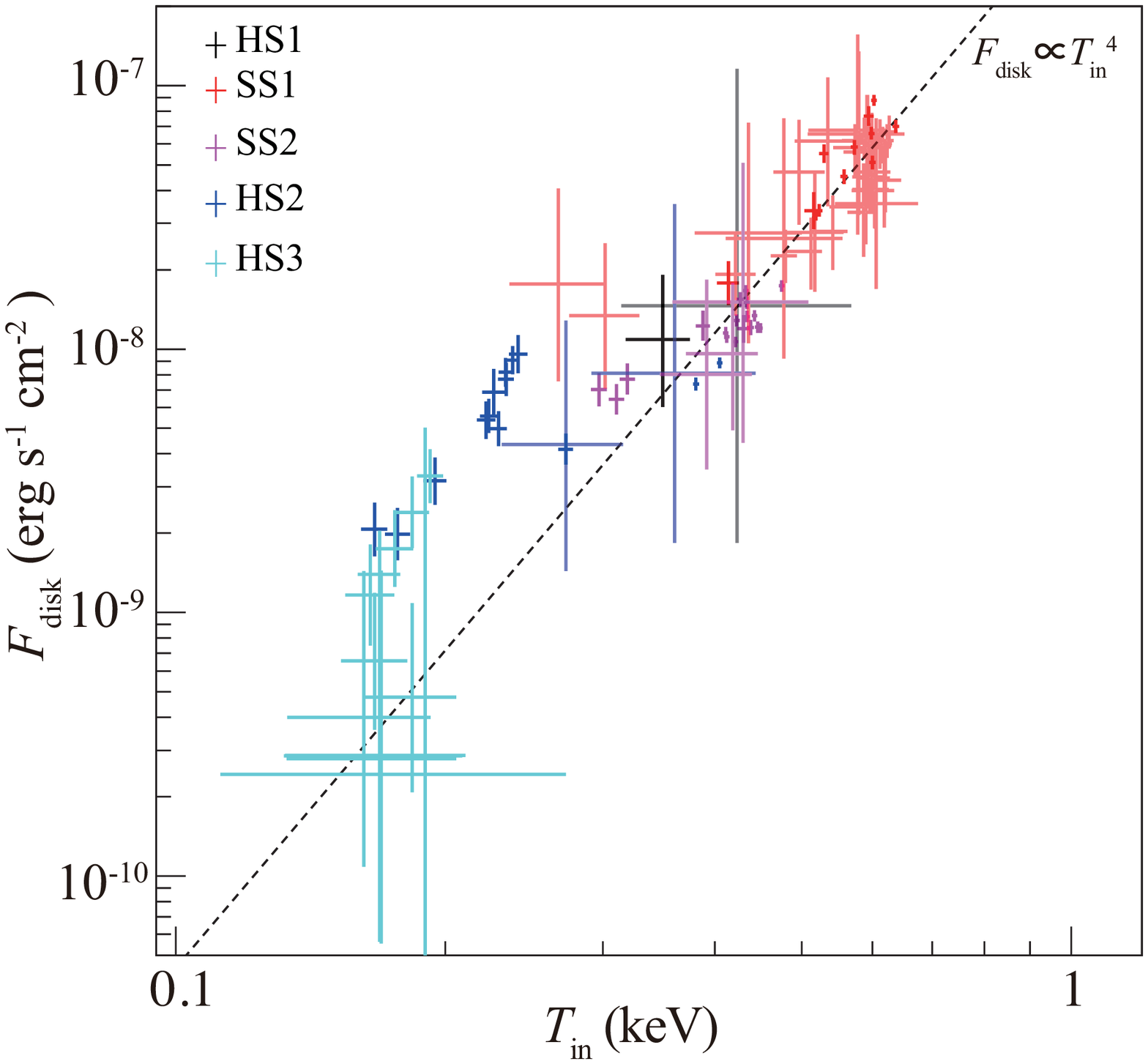}
  \end{center}
  \caption{
  The correlation between $\Tin$ and Comptonization-corrected disk flux, 
  data points in each periods are shown in the same colors as in figure \ref{fig:SwiftHID} and \ref{fig:FLUX_UVM2}.
  The data taken by {\swift}/XRT are shown in solid colors, and those by {\maxi}/GSC in faint colors.
  The dashed line indicates $\fdisk\propto$ $\Tin^4$.}
\label{fig:Tin_DISKFLUX}
\end{figure}

The MCD component was significantly detected in both the SS and the HS in the {\swift}/XRT snapshot spectra.
In the SS, the innermost disk radius was almost constant at around 150 km despite a large change in flux 
over an order of magnitude,
while $\rin$=300--400 km was obtained in the HS.
We have shown the $\rin$ versus disk-flux $\fdisk$ diagram in figure \ref{fig:Tin_DISKFLUX}.
In the SS, the relation between $\fdisk$, $\Tin$ and $\rin$ is known to be $F_{\rm disk} \propto \Tin^4 \rin^2$.
During the SS1 and the SS2, $\fdisk$ tracks on $F_{\rm disk} \propto \Tin^4$.
However, it clearly deviates in the HS2 and the first half of the HS3,
implying the truncation of the inner disk \citep{truncate}.

\begin{figure}
  \begin{center}
    \FigureFile(90mm,90mm){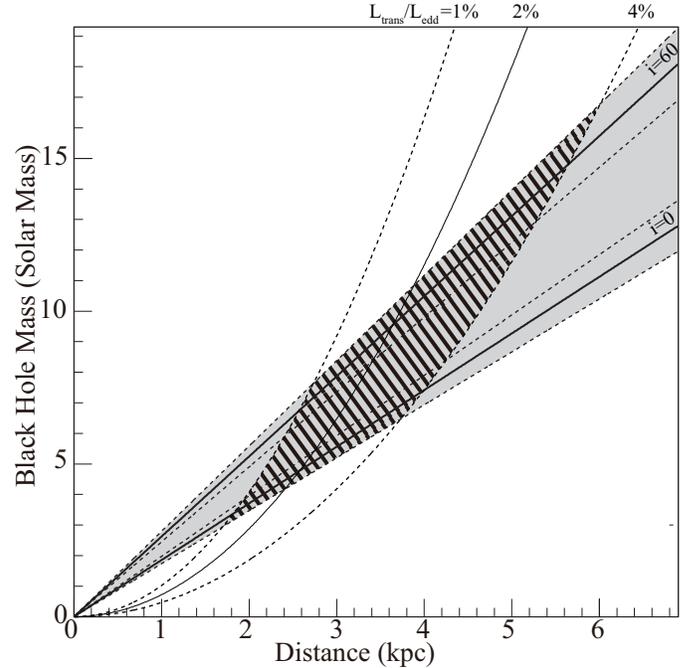}
  \end{center}
  \caption{ 
   Observational constraints on the distance-mass diagram.
   The shadowed region comes from the innermost radius derived from X-ray spectral analysis 
   assuming an inclination angle of 0 to 60 degree, and the solid lines with the two dashed 
   lines in the region indicate the best-fit parameters and their errors when $i=0^\circ$ and $60^\circ$.
   The hatched area is derived by further assuming the empirical relation that 
   the soft to hard transition occurs at 1--4\% $\ledd$.
  }
\label{fig:MASS_DIST}
\end{figure}

Next, we estimate the mass of the central star, firstly assumed to be a black hole, from the inner disk radius.
If a color correction factor $\kappa$ 
has little dependence of  $r$ \citep{mcd_hardning}, 
the innermost radius $\Rin$ in the standard accretion disk model 
can be approximately expressed by $\Rin=\kappa^2 \xi \rin$, where $\xi$ is a correction factor for 
a maximal flux or temperature radius.
Though both $\kappa$ and $\xi$ have some uncertainties 
(e.g., see \cite{davis_spec} for $\kappa$, and see \cite{dotani_mcd} for $\xi$), 
we take $\kappa = 1.7$ and $\xi = 0.412$  \citep{true_Rin} to directly compare with our previous results
(\cite{nkhr1752b}, \cite{1659yama}, \cite{1305mrhn}).
By adopting the SS2 averaged radius of $138.0\pm9.0$ km as ISCO, 
the mass for a non-spinning black hole is estimated as
\begin {eqnarray}
\label{eq-mass}
M_{BH} = \frac{c^2\Rin}{6G} = 18.5\pm1.2\left(\frac{D}{10 {\rm kpc}}\right) (\cos{\it i})^{-\frac{1}{2}} M_{\odot} .
\end {eqnarray}
with a source distance in unit of 10 kpc and an inclination angle from our line-of-sight $i$.
Since no dip or eclipse feature was observed, we assume $i = 0-60^\circ$.
Thus, the distance and mass is constrained to the shadowed region in figure \ref{fig:MASS_DIST}.
Following \citet{nkhr1752b} and \citet{1659yama}, we derive further constraints 
using an empirical correlation offered by \citet{ltrans} that the soft-to-hard 
spectral state transition occurs between 1 to 4 \% of Eddington luminosity.
If we assume that the soft-to-hard transition fluxes of 1.28--2.33 $\ergcmse$ obtained 
in section \ref{xray_spec} lie in this $\ledd$ range,
then the hatched region in figure \ref{fig:MASS_DIST} can be derived, 
and the minimum and maximum masses are obtained as
2.9 $\msolar$ for $D=1.7$ kpc and 12.9 $\msolar$ for $D=5.3$ kpc, respectively.
The lower limit close to 3 $\msolar$ is consistent with a picture that the central star is a black hole 
as assumed in the mass derivation.
A further detailed estimation of the mass fully taking account of the general  relativistic effects is 
out of the scope of this paper.

We note that $\rin$ used in the SS2 is slightly different from that in the SS1, which might be due to 
that Compton scattering above the disk is partially expressed by the {\em nthcomp} model, 
not the hardening factor $\kappa$. We also note that a larger radius $\rin \simeq 200$ km was 
obtained on $t \sim140$ d,  when {\em XMM-Newton} observation was carried out \citep{1910reis}.
The larger radius than that in the SS implies the truncation of the optically thick disk in that state,
rather than an intrinsically large innermost radius of a retrogradely spinning black hole \citep{1910reis}.

\subsection{Origin of UV and X-ray Dip and Re-flareing}

The ``fast-rise and exponential-decay'' type of outbursts is well explained by a sudden and temporary increase 
in mass accretion induced, for instance, by a disk thermal instability (\cite{mineDTI}, \cite{DTI}).
During the steady-decline phases in $t$ = 10--90 d and 110--140 d,
both the $\fx$ and $\fuv$ light curves of MAXI~J1910--057 exhibited exponential decays  (figure \ref{fig:CCF}).
Behaviors of the curves around $t$ = 90--110 d and after $t$ = 140 d are, however, different.
In $t=90-110$ d, both the curves are characterized by the sharp drop, and $\fx$ lags behind $\fuv$.
We computed the time delay and the profile focusing on the sharp drop in section 3.4.
By employing the convolution model with a Gaussian kernel, we obtained 
the lag time of 3.5$\pm0.2$ days and the diffusion time of 3.8$_{-0.7}^{+0.8}$ days (FHWM).
The lag time is consistent with $4.06\pm0.71$ days obtained by \citet{1910Degenaar} who used the same Swift UVOT and XRT data.
We also note that the hysteresis in the right-top of figure \ref{fig:FLUX_UVM2} 
suggests the same delay in the early part of the outburst.

What does the dip mean?
\citet{1910Degenaar} extensively argued the cause of the dip in connection with jets and state transitions.
Here, we also discuss a possibility of non-steady accretion flow. 
In the optically thick accretions disk, the disk temperature $T$ decreases outward as $T \propto r^{-3/4}$.
Assuming most photons having energy $E$ come from the region with temperature $T (\sim E/k)$
according to the Wien's displacement law, we observe photons with $E$  from the radius at $r \propto E^{-4/3}$.
Thus, it is possible that the observed energy and time dependent dip is caused by a low mass-accretion-rate (dark ring) region
taking place at an outer part of the disk and drifting inward.
From the SED at $t=93.3$ d (figure \ref{fig:uvot_sed}(d)), 
a ratio of the X-ray-flux-peak energy $E_{\rm X}$ to the UVOT (filter) central energy (2246 \AA) $E_{\rm UV}$, 
$E_{\rm X}/E_{\rm UV}$,  is $\sim 100$.
Thus, the radius of the UV emission region $r_{\rm UV}$  can be roughly estimated as 
$r_{\rm UV} \simeq (E_{\rm UV}/E_{\rm X})^{-4/3}r_{\rm in} \sim500\ r_{\rm in} \sim 3000\  r_{\rm s}$.
Here we assume that X-rays with $E_{\rm X}$ mostly come from near $r_{\rm in}$, and 
use $r_{\rm in} \simeq  2R_{\rm in} (\simeq 6\ r_{\rm s})$ from Tables \ref{table:specana} or \ref{table:uvspecana}.
We ignore the effect of the irradiation here.

This UV radiation radius is comparable to the radius estimated from the viscous time scale for 6 days 
of Optical-to-X-rays time lags \citep{1910Degenaar} though the viscous time scale has 
the uncertainty of one or two orders of magnitude.
It is interesting to note that the viscous time scale is also known as the (radial) {\it drift} time scale \citep{lightman74} 
and the {\it diffusion} time scale \citep{accpower}.
Thus, the time lags ($\sim 3.5$ days) and the comparable diffusion time ($\sim 3.8$ days) we obtained are consistent
with the above picture. 

The dip might be also associated with the bright 3rd maximum around $t\sim120$ d
because a similar X-ray flux drop before the 3rd maximum was also observed in A~0620-00 \citep{0620}.
An apparent flux excess over the exponential decay implies enhancement of a mass-accretion rate 
by an extra mass input to the disk. 
If the accretion rate decreases at an outer part of the disk, 
a transition into a cool ($\sim 6000$ K) branch of a Hydrogen neutral state sets in \citep{dwarfnova}.
In the cool branch, the viscosity is so small that the mass accretion once stagnates there.
This might result in the dark ring region just inside it. 
The transition also causes a vertical shrink of the outer disk, allowing  irradiation to the companion star.
This gives rise to an additional mass transfer to the accretion disk invoked by mass transfer instability  (e.g., \cite{2ndmaxchen}),
and the disk backs to a hot branch of a Hydrogen ionized state.

Finally, we also note that the UV dip might be partially triggered by the transition because the temperature of the hot branch
is about $10^4$ K, of which the peak energy is around 3.4 eV.
As pointed out by \citet{1910Degenaar}, however, these scenarios can not explain the sudden nIR color change (bluer) 
observed with the sharp drop in UVs.

After $t=140$ d, $\fx$ and $\fuv$ changed asymmetrically. 
From the fits to the Swift/XRT plus UVOT spectra with the reprocessed disk model (figure \ref{fig:uvot_sed}), 
a reprocessed fraction can be considered to be increasing with time.
In the present data, however, we cannot rule out other possibilities. 
The UV component might be due to other components such as a jet and/or an inner hot accretion flow (e.g., \cite{yuanjet}), 
and the soft X-ray component also to an ``additional variable component''  \citep{additional_comp}, 
e.g., X-ray shots with a soft X-ray spectrum and associated optical jets (\cite{ngr_shot94}, \cite{malzac03}), 
instead of the Keplerian (less variable) disk component.

\section{Conclusion}

We have presented light curves and spectral results of the entire outburst 
using data obtained by {\maxi} and {\swift} of the X-ray nova MAXI\,J1910--057.
The light curve was a fast-rise and exponential-decay type suggesting 
a disk thermal instability outburst. 
As well as past observations of X-ray novae, the broad third maximum was detected. 
The X-ray spectral state around the main peak and the third maximum were 
confirmed to be the SS.
Since the innermost radius kept roughly constant in the SS, 
and the relation between disk flux and temperature was approximated 
by $\fdisk \propto \Tin^4$, we interpreted the radius as an ISCO.
The innermost radius became larger in the hard state. 
This suggests that the accretion disk was truncated.
The X-ray fluxes of two series of ``hard-to-soft'' transitions agreed
within a factor of two.
By assuming that the ``hard-to-soft'' transitions occurs 1--4\% of Eddington luminosity, 
the lower limit of the black hole mass is estimated to be 2.9 $\msolar$.

We also studied the relation between UV data in combination with X-ray data.
Their joint spectral analysis revealed that the UV emission is composed of
direct MCD emission and its reprocessed emission from outer disk.
Using the sharp drop found in both the light curves, the profile of time delay 
is approximated with a Gaussian profile centering at 3.5 days 
and the width of 3.8 days FWHM.
We conclude that the time difference reflects changing mass 
accretion rate accreting inward, which possibly caused by local instability.

This research has made use of {\maxi} data provided by RIKEN, JAXA and
the {\maxi} team. We also thank the {\swift} team for their observation. 
This work made use of data supplied by the UK {\swift} Science Data Centre at 
the University of Leicester.


\end{document}